\documentclass[12pt,headings=big,numbers=noenddot,DIV=14,a4paper]{article}%

\pdfoutput=1

\usepackage[margin=1 in]{geometry}

\usepackage{graphicx}  
\usepackage{dcolumn}   
\usepackage{bm,relsize}        
\usepackage{amsfonts,amsmath,amssymb,amsthm,mathtools}
\usepackage{slashed} 
\usepackage[normalem]{ulem}
\usepackage{placeins}
\usepackage{afterpage}
\usepackage{lipsum, color}
\usepackage[usenames,dvipsnames,svgnames,table]{xcolor}
\usepackage[linktoc=page,bookmarks=false,colorlinks=true,linkbordercolor=RoyalBlue,citebordercolor=ForestGreen,urlbordercolor=CornflowerBlue]{hyperref}
\hypersetup{
    colorlinks=true,
    linkcolor=ForestGreen,
    filecolor=magenta,      
    urlcolor=ForestGreen,
    citecolor=ForestGreen,
}
\usepackage[sort&compress,numbers,merge]{natbib}

\def\beq{\begin{equation}}
\def\eeq{\end{equation}}
\def\bea{\begin{eqnarray}}
\def\eea{\end{eqnarray}}

\def\tauV		{\tau_{\mathsmaller{V}}}
\def\mV			{m_{\mathsmaller{V}}}
\def\MDM		{M_{\mathsmaller{\rm DM}}}

\def\sSM		{s_{\mathsmaller{\rm SM}}}

\def\TD			{T_{\mathsmaller{D}}}
\def\TSM		{T_{\mathsmaller{\rm SM}}}

\def\rtilde		{\tilde{r}}
\def\gtildeD	{\tilde{g}_{\mathsmaller{D}}}
\def\gtildeSM	{\tilde{g}_{\mathsmaller{\rm SM}}}

\begin{document}

\begin{flushright}
\footnotesize
DESY 19-206\\
\end{flushright}
\color{black}

\begin{center}

{\LARGE \bf BSM with Cosmic Strings: Heavy, up to EeV mass, Unstable Particles\\
\vspace{.3 cm}
\large
}

\medskip
\bigskip\color{black}\vspace{0.5cm}

{
{\large Yann Gouttenoire}$^a$,
{\large G\'eraldine Servant}$^{a,b}$,
{\large Peera Simakachorn}$^{a,b}$
}
\\[7mm]

{\it \small $^a$ DESY, Notkestra{\ss}e 85, D-22607 Hamburg, Germany}\\
\it \small $^b$ II. Institute of Theoretical Physics, Universit\"at Hamburg, 22761 Hamburg, Germany \\
\end{center}

\bigskip

\centerline{\bf Abstract}
\begin{quote}
\color{black}

Unstable heavy particles well above the TeV scale are unaccessible experimentally.
So far, Big-Bang Nucleosynthesis (BBN) provides the strongest limits on their mass and lifetime, the latter being shorter than 0.1 second.
We show how these constraints could be potentially tremendously improved by the next generation of Gravitational-Wave (GW) interferometers, extending to lifetimes as short as $10^{-16}$ second.
The key point is that these particles
may have dominated the energy density of the universe and have triggered a period of matter domination at early times, until their decay before BBN.
The resulting modified cosmological history compared to the usually-assumed single radiation era 
would imprint observable signatures in stochastic gravitational-wave backgrounds of primordial origin.
In particular, we show how the detection of the GW spectrum produced by long-lasting sources such as cosmic strings would provide a unique probe of particle physics parameters. 
When applied to specific particle production mechanisms in the early universe, these GW spectra could be used to derive new constraints on many UV extensions of the Standard Model. We illustrate this on a few examples, such as supersymmetric models where the mass scale of scalar moduli and  gravitino can be constrained up to $10^{10}$~GeV. Further bounds can be obtained on the reheating temperature of models with  only-gravitationally-interacting particles as well as on the kinetic mixing of  heavy dark photons at the level of $10^{-18}$.

\end{quote}

\clearpage

\tableofcontents


\section{Introduction}

The existence of very massive particles $X$, with mass $m_X \gg $ TeV,  is a generic prediction of many well-motivated extensions  of the Standard Model (SM) of particle physics, such as Grand Unified Theories, extra-dimensional models inspired by String Theory or supersymmetric constructions.
If such particles are stable and still present in our universe today, 
they can contribute to the dark matter, in which case 
a variety of detection strategies has been explored depending on their mass range and the nature of their interactions. On the other hand, unstable particles beyond the Standard Model (BSM) are very difficult to probe experimentally. The best chances are through their effects on cosmological  observables. The strongest limits come from Big-Bang Nucleosynthesis (BBN),  since any heavy relic which decays after BBN would ruin the predicted abundances of light elements.  From BBN, one obtains general model-independent bounds in the plane $(\tau_X,\, m_X Y_X)$ where $\tau_X$ is their lifetime, $m_X$ their mass and $m_X Y_X$ is their would-be contribution to the total energy density of the universe today if they had not decayed \cite{Cyburt:2002uv,Jedamzik:2006xz, Jedamzik:2009uy, Kawasaki:2017bqm}. We can therefore infer indirect information on their couplings through the constraints on their lifetime and the efficiency of their production mechanism in the early universe.
In the present work, we show how a large new region unexplored so far in the  $(\tau_X,m_XY_X)$  plane can be probed using future gravitational-wave observatories.

Our starting assumption is that these particles can temporarily dominate the energy density of the universe, and therefore induce a period of matter domination within the radiation era after post-inflationary reheating. This leads to a modified expansion of the universe compared to the usually assumed single radiation era. Interestingly, such modified cosmological history can be probed if during this period, there is an active source of gravitational waves, in which case the resulting GW spectrum would imprint any modification of the equation of state of the universe. 
Particularly well-motivated are the long-lasting GW production from cosmic string networks.

Cosmic strings (CS) are topological defects produced when a $U(1)$ symmetry gets spontaneously broken in the early universe \cite{Hindmarsh:1994re, Vilenkin:2000jqa, Vachaspati:2015cma}. 
CS behave as dynamical classical objects, moving at relativistic speed. 
They can also be described by fundamental or composite objects in string theory \cite{Witten:1985fp, Dvali:2003zj,Copeland:2003bj, Polchinski:2004ia, Sakellariadou:2008ie, Davis:2008dj, Sakellariadou:2009ev, Copeland:2009ga}.
CS networks offer promising prospects for the detection of GW of cosmological origin. They are a generic prediction of many Standard Model extensions, such as models of Grand Unification \cite{Jeannerot:2003qv,Sakellariadou:2007bv,Buchmuller:2013lra}, or the seesaw mechanism for neutrino masses when $U_{B-L}(1)$ is broken spontaneously \cite{Dror:2019syi}.

The string network is characterized by its correlation length $L$. While strings are stretched by cosmic expansion, they form loops. One would naively expect $L$ to evolve linearly with the scale factor $a$ due to the Hubble expansion, such that  $L\propto t^{1/2}$ in radiation domination and $L \propto t^{2/3}$ is matter domination. However, a remarkable feature of CS is that, after a transient evolution, the system reaches its  scaling regime where the energy loss of long strings into loop formation, is exactly such that $L$ scales linearly with the Hubble horizon $t$ \cite{Ringeval:2005kr, Vanchurin:2005pa, Martins:2005es, Olum:2006ix, BlancoPillado:2011dq}. During this regime, the string network is only characterized by the string tension $\mu$, roughly given by the square of the phase transition temperature $T_{p}$
\begin{equation}
G\mu \simeq 10^{-15} \left( \frac{T_{p}}{10^{11}~\text{GeV}} \right)^2,
\end{equation}
and the long string density, $\rho_{\infty} = \mu/L^2$, redshifts as radiation in radiation domination and as matter in matter domination. 

It has since long been conjectured that the oscillations of the CS loops may be the dominant source of the Stochastic Gravitational Waves Background (SGWB), principally because they constitute a long-standing source starting GW emission after the network formation and still radiating today \cite{Vilenkin:1981bx, Hogan:1984is, Vachaspati:1984gt, Accetta:1988bg, Bennett:1990ry, Caldwell:1991jj,Allen:1991bk, Battye:1997ji, DePies:2007bm, Siemens:2006yp, Olmez:2010bi, Regimbau:2011bm, Sanidas:2012ee, Sanidas:2012tf, Binetruy:2012ze, Kuroyanagi:2012wm, Kuroyanagi:2012jf}. The interesting aspect of the SGWB from CS is its flat frequency spectrum, assuming standard cosmology, spanning many orders of magnitude in frequency. Hence, the capability of the next generation of GW interferometers, LISA \cite{Audley:2017drz}, Einstein Telescope \cite{Hild:2010id, Punturo:2010zz}, Cosmic Explorer \cite{Evans:2016mbw}, BBO and DECIGO \cite{Yagi:2011wg} to detect the SGWB from CS, opens a unique observational window, on any new physics likely to change the thermal history of the universe, and imprint features in the GW spectrum. In \cite{CS_VOS}, we analysed in detail how much information on the early  universe equation of state could be extracted from the observation of a SGWB generated by CS, adding up on  recent works   \cite{Cui:2017ufi, Cui:2018rwi, Auclair:2019wcv, Guedes:2018afo}.

In this study, we assume the presence of a heavy, cold, particle $X$ dominating the energy density of the universe at the temperature $T_{\rm dom}$ and decaying at the temperature $T_{\rm dec}$. Currently, the success of BBN in a standard radiation dominated universe provides the strongest constraint on such scenario, $T_{\rm dec} \gtrsim 1$~MeV.
The key point of our study is that the observation of a flat GW spectrum from CS would extend by far the BBN constraints on heavy relics.
All assumptions relevant for our conclusions are discussed in detail in our companion article \cite{CS_VOS}, where we refine and extend the work of \cite{Cui:2017ufi, Cui:2018rwi} beyond the scaling regime during the change of cosmology. We also take into account the recent discussion on the effect of  particle production on the GW emission \cite{Auclair:2019jip}. We only recap  briefly the main points here and refer the reader to \cite{CS_VOS} for more details.

In Sec.~\ref{sec:recap}, we  review the main assumptions about the microscopic description of CS.
In Sec.~\ref{sec:NSmatter}, we compute the GW spectrum in the presence of an early matter era. We then derive the improvement by many orders of magnitude of the current model-independent BBN constraints  on the abundance and lifetime of a particle, c.f. Fig.~\ref{fig:mY_tauX_GWI_VS_BBN}, that can be inferred from the detection of GW produced by CS. In Sec.~\ref{sec:BenchmarkModels}, we provide unprecedented exclusion bounds on four particle physics models leading to an early matter domination era: oscillating scalar moduli in supersymmetric theories, {secluded scalar sectors which are only gravitationally produced}, scalars produced through the Higgs portal, and massive dark photons. {At the very end, we also study the scenario where the dark photon mass and the cosmic string network are generated by the spontaneous breaking of the same $U(1)$ symmetry. }
\section{Gravitational waves from cosmic strings}
\label{sec:recap}
\subsection{The micro-physics assumptions}

A detailed discussion of the assumptions made for the prediction of SGWB from cosmic strings is provided in 
our companion article  \cite{CS_VOS}.  We summarize them here. 

\begin{itemize}
\item We consider CS from the breaking of a gauge symmetry.

\item We assume first that CS can be described by Nambu-Goto strings which are 1-dimensional classical object, characterized solely by their tension $\mu$. The main decay channel of loops are through gravitational waves and their sizes shrink with a rate $\Gamma G \mu$ with $\Gamma \simeq 50$ \cite{Blanco-Pillado:2017oxo}. 

\item As a correction to the Nambu Goto approximation, we include the possibility of producing massive particles due to the presence of string segments with curvature comparable to their thickness, cusps and kinks.

\item We consider a monochromatic loop distribution, where loops are produced with a length being a fraction $\alpha=0.1$ of the Hubble size. Hence, we neglect the contributions to the SGWB from the loops produced at smaller scales. In particular, we neglect the contributions from the loops produced at the gravitational back reaction scale, the scale below which small-structures are smoothened.

\item GW production at time $\tilde{t}$ is dominated by the emission from small loops of size $\sim \Gamma G \mu \, \tilde{t}$, which have been produced at the horizon size, i.e. much earlier, at time $t_i \sim \Gamma G \mu \,\tilde{t}/\alpha$.

\item We compute the loop-formation efficiency $C_{\rm eff}$ by solving the Velocity-dependent One-Scale equations  
\cite{Kibble:1984hp,Martins:2000cs,Martins:1995tg,Martins:1996jp,martins2016defect}. 
Hence, we account for the extra time needed by the long-string network to respond to the change of cosmology, inducing a shift by one or two orders of magnitude of the turning-point frequency characterizing the non-standard matter era.

\end{itemize}

\subsection{The gravitational-wave spectrum}
For our study, we follow the assumptions listed above
and the expression given in \cite{Cui:2018rwi} for the computation of today GW spectrum generated from CS (for a derivation and discussion of this formula, see \cite{CS_VOS})
	\begin{equation}
	\Omega_{\rm{GW}}(f)\equiv\frac{f}{\rho_c}\left|\frac{d\rho_{\rm{GW}}}{df}\right|=\sum_k{\Omega^{(k)}_{\rm{GW}}(f)},
	\label{eq:spect_master}
	\end{equation}
	with
	\begin{equation}
\Omega^{(k)}_{\rm{GW}}(f)=\frac{1}{\rho_c}\cdot\frac{2k}{f}\cdot\frac{(0.1) \,\Gamma^{(k)}G\mu^2}{\alpha(\alpha+\Gamma G \mu)}\int^{t_0}_{t_F}d\tilde{t} \, \frac{C_{\rm{eff}}(t_i)}{t_i^4}\left[\frac{a(\tilde{t})}{a(t_0)}\right]^5\left[\frac{a(t_i)}{a(\tilde{t})}\right]^3\Theta(t_i-t_F)\Theta(t_i-\frac{l_*}{\alpha}),
	\label{kmode_omega}
	\end{equation}
where we integrate over all emission times $\tilde{t}$ starting from the network formation time $t_F$ up to today $t_0$ and where the sum is performed over the mode frequencies of the loop.
We assume that a loop created at $t_i$, with a length at a fraction $\alpha$ of the Hubble horizon, shrinks by emission of GW with a rate $\Gamma G \mu$
\begin{equation}
\label{eq:CSlength}
l(t) = \alpha t_i -\Gamma G \mu(t-t_i),
\end{equation}
where the gravitational loop-emission efficiency is independent of the lenght of the loop, $\Gamma \simeq 50$ \cite{Blanco-Pillado:2017oxo}. The 1-loop power spectrum $\Gamma^{(k)}\, G \mu^2$ assumes cusp domination
\begin{equation}
\Gamma^{(k)}= \frac{\Gamma \, k^{-4/3} }{ \sum_{p=1}^{\infty}p^{-4/3}} \simeq \frac{\Gamma \, k^{-4/3} }{ 3.60}.
\end{equation}
From Eq.~\eqref{eq:CSlength}, the time $t_i$, when the loops, sourcing GW emitted at time $\tilde{t}$ and detected with frequency today $f$, have been created can be expressed as
\begin{equation}
\label{eq:def_t_i}
t_i (f, \, \tilde{t})= \frac{1}{\alpha+\Gamma G \mu} \left[ \frac{2k}{f}\frac{a(\tilde{t})}{a(t_0)} + \Gamma G \mu \, \tilde{t} \right].
\end{equation}
As discussed in \cite{Auclair:2019jip, CS_VOS}, loops with length smaller than $l_*$ decay dominantly into massive particles
\begin{align}
l_*=\beta_m\frac{\mu^{-1/2}}{(\Gamma G\mu)^m},
\end{align}
where $m=1$ or $2$ for kink-dominated or cusp-dominated loops, respectively, and $\beta_m\sim\mathcal{O}(1)$.

\begin{figure}[h!]
\centering
\raisebox{0cm}{\makebox{\includegraphics[height=0.55\textwidth, scale=1.]{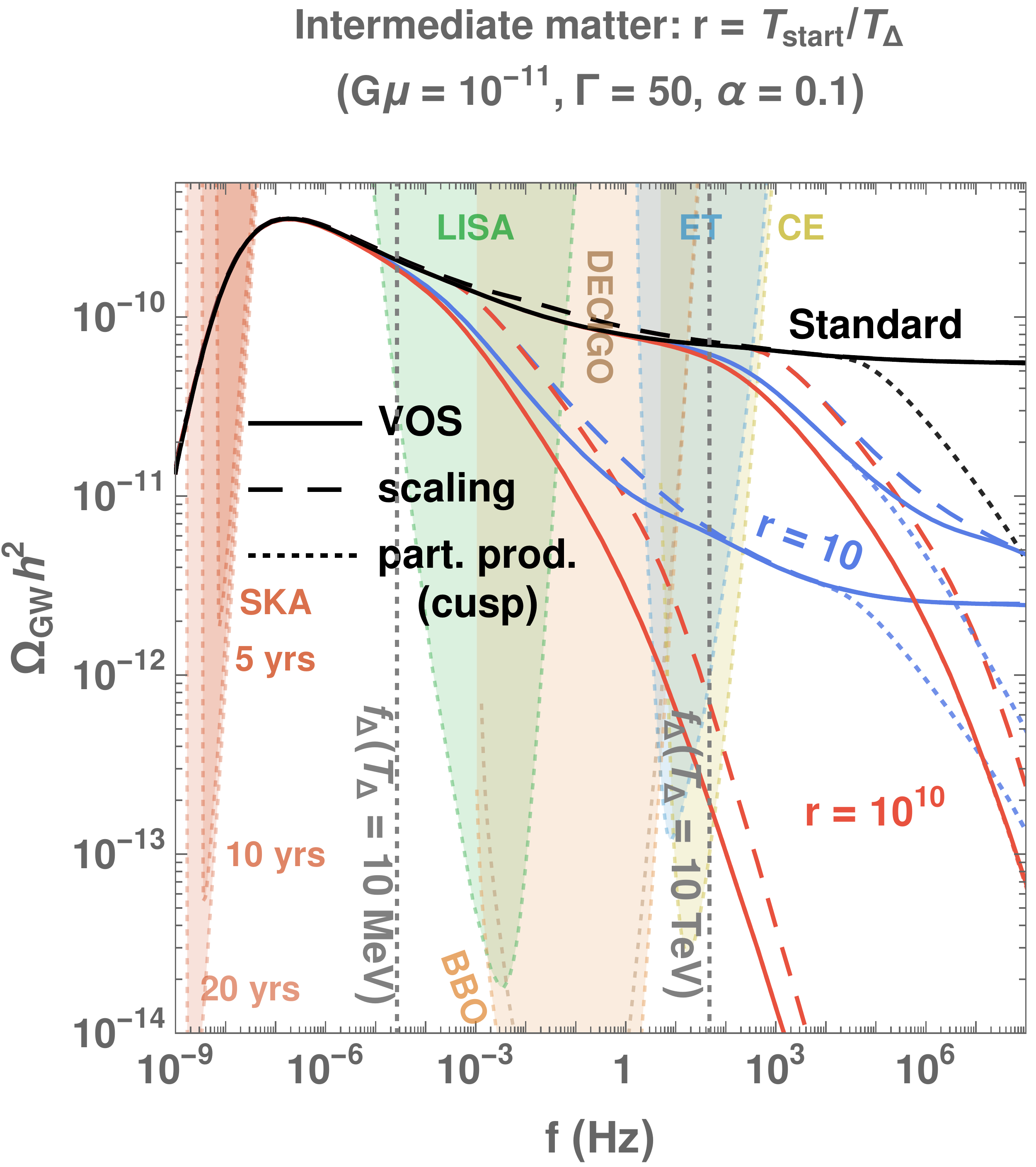}}}
\raisebox{0cm}{\makebox{\includegraphics[height=0.55\textwidth, scale=1.]{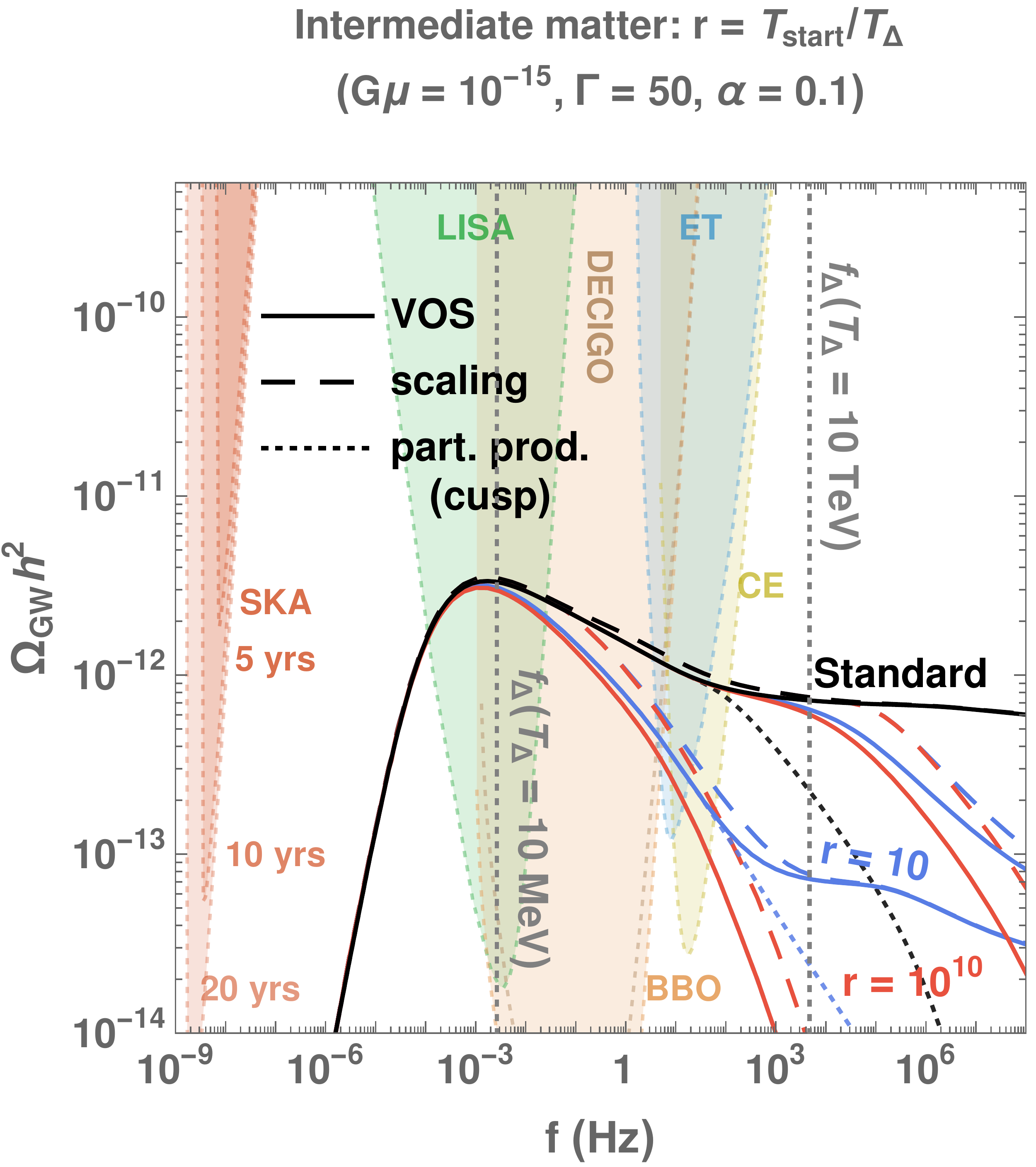}}}
\caption{\it \small \label{fig:OmegaGWCS}  SGWB generated by the gravitational decay of cosmic strings compared to the reach of different GW interferometers. We show the impact of a long (red) or a short (blue) intermediate matter era, starting at the temperature $r \, T_{\Delta}$ and ending at $T_{\Delta} = 10$~MeV or $T_{\Delta} = 10$~TeV. Black lines show the results obtained assuming  standard cosmological evolution. The dashed-lines assume that the scaling regime switches on instantaneously during the change of cosmology whereas the solid lines incorporate the transient behavior, solution of the VOS equations, as discussed in \cite{CS_VOS}. 
Limitations due to particle production assuming that the small-scale structures are dominated by cusps are shown with dotted lines \cite{CS_VOS}.
The dotted vertical lines indicate the relation in Eq.~\eqref{eq:f_delta_RD} between the temperature $T_{\Delta}$ and the frequency $f_{\Delta}$ of the turning point, where the matter-era-tilted spectrum meets the radiation-era-flat spectrum.  }
\end{figure}
\subsection{The Velocity-dependent One-Scale model}
Numerical simulations realized by Blanco-Pillado et al. \cite{Blanco-Pillado:2013qja} have shown that a fraction  $\mathcal{F}\simeq 10\%$ of the loops are produced with a length equal to $\alpha \simeq 10\%$ of the horizon size and with a Lorentz boost factor $\gamma \simeq \sqrt{2}$. The remaining $90\%$ of the energy loss by long strings goes into highly boosted smaller loops whose contribution to the GW spectrum is sub-dominant. Under those assumptions, the Velocity-dependent One-Scale (VOS) model \cite{Martins:1995tg, Martins:1996jp, Martins:2000cs, martins2016defect} predicts the loop-formation rate to be
\begin{equation}
\label{eq:LoopProductionFctBody}
\frac{dn}{dt_i}=\mathcal{F} \frac{C_{\rm eff}(t_i)}{\alpha \, t_i^4}.
\end{equation}
Concretely, the VOS model describes the evolution of Nambu-Goto long strings through two macroscopic quantities, their mean velocity $\bar{v}$ and correlation length $\xi \equiv L/t$, from which one can compute the loop-formation efficiency
\begin{equation}
\label{eq:LoopFormationEfficiency}
C_{\rm eff} \equiv \frac{\tilde{c}\,\bar{v}}{\sqrt{2}\xi^3},
\end{equation}
where $\tilde{c}$ is the loop-chopping efficiency, $\tilde{c}=0.23$ based on Nambu-Goto numerical simulations \cite{Martins:2000cs}. In \cite{CS_VOS}, we study the implication of the recent refinement of the VOS model \cite{Correia:2019bdl}, which includes particle production and is based on abelian-Higgs field theory numerical simulations.
Out of the competing dynamics of Hubble stretching and energy loss through loop chopping, the network reaches an attractor solution, the \textit{scaling regime}, in which both $\bar{v}$ and $\xi$ are \textit{constant}. In our companion paper \cite{CS_VOS},  we show that the long-string network deviates from the scaling regime during a change of cosmology. As an illustration, Fig.~\ref{fig:OmegaGWCS} compares the spectrum computed by assuming that the scaling regime is reached instantaneously during the change of cosmology with the full solution of the VOS equations. It shows that the turning point frequency which is a signature of the change of cosmology from matter to radiation, is over-estimated by more than one order of magnitude in the scaling approximation. Ref.~\cite{CS_VOS} refines  the study of \cite{Cui:2018rwi} by going beyond the scaling regime as well as considering particle production and additional non-standard cosmological histories.

\subsection{The reach of GW interferometers}

GW spectra from CS for two values of $G\mu=10^{-15}$, $10^{-11}$ are plotted in Fig.~\ref{fig:OmegaGWCS}, together with the power-law sensitivity curves of NANOGrav \cite{Arzoumanian:2018saf}, EPTA \cite{vanHaasteren:2011ni}, SKA \cite{Janssen:2014dka}, LIGO \cite{Aasi:2014mqd}, DECIGO, BBO \cite{Yagi:2011wg}, LISA \cite{Audley:2017drz}, Einstein Telescope \cite{Hild:2010id, Punturo:2010zz} and Cosmic Explorer \cite{Evans:2016mbw}. Only EPTA, NANOGrav and LIGO O1/O2 (which are not visible on the plots) are current constraints, the other ones being projected sensitivities of future projects. The power-law integrated sensitivity curves are computed in \cite{CS_VOS} with a signal-to-noise ratio SNR=$10$ and an observation time of $10~$years. At lower string tension $G\mu$, the GW power emission per loop is smaller, hence the amplitude is suppressed. Also, for lower values of $G\mu$, loops decay more slowly and GW are emitted later, which implies a lower red-shift factor and a global shift of the spectrum towards higher frequency.

The strongest constraints come from pulsar timing array EPTA, $G \mu \lesssim 8 \times 10^{-10}$ \cite{Lentati:2015qwp}, and NANOGrav,  $G \mu \lesssim 5.3 \times 10^{-11}$ \cite{Arzoumanian:2018saf}, therefore we limit ourselves to $G\mu < 10^{-11}$. 
Our analysis is based on the assumption that the astrophysical foreground can be subtracted.
The GW spectrum generated by the astrophysical foreground increases with frequency as $f^{2/3}$ \cite{Zhu:2012xw},  differently from the GW spectrum generated by CS during radiation (flat) or during matter ($f^{-1}$).

\section{The imprints of an early era of matter domination}
\label{sec:NSmatter}

\subsection{Modified spectral index}

The part of the spectrum coming from loops produced and emitting during radiation is flat since there is an exact cancellation between the red-tilted red-shift factor and the blue-tilted loop number density. However, in the case of a matter era, a mismatch induces a slope $f^{-1}$. The impact of a non-standard matter era is shown in Fig.~\ref{fig:OmegaGWCS}. The frequency detected today $f_{\Delta}$ of the turning point between the end of the matter domination and the beginning of the radiation-domination can be related to the temperature of the universe $T_\Delta$ when the change of cosmology occurs 
\begin{equation}
\label{eq:f_delta_RD}
f_{\Delta}=                  (2 \times 10^{-3} ~\textrm{Hz})\left( \frac{T_{\Delta}}{\textrm{GeV}} \right) \left( \frac{0.1 \times 50 \times 10^{-11}}{\alpha \Gamma G_{\mu}} \right)^{1/2} \left( \frac{g_{*}(T_{\Delta})}{g_{*}(T_{0})}\right)^{1/4}. 
\end{equation}
The GW  measured with frequency $f_{\Delta}$ have been emitted by loops produced during the change of cosmology at $T_{\Delta}$.
An extensive discussion of this frequency-temperature relation as provided  in \cite{CS_VOS}. 
The above formula  entirely relies on the assumptions that the back-reaction scale is $\Gamma G \mu$ as claimed by Blanco-Pillado et al. \cite{Lorenz:2010sm, Ringeval:2017eww, Auclair:2019zoz} and not much lower as claimed by Ringeval et al. \cite{Blanco-Pillado:2013qja, Blanco-Pillado:2017oxo} . 

\begin{figure}[h!]
\centering
\raisebox{0cm}{\makebox{\includegraphics[height=0.52\textwidth, scale=1]{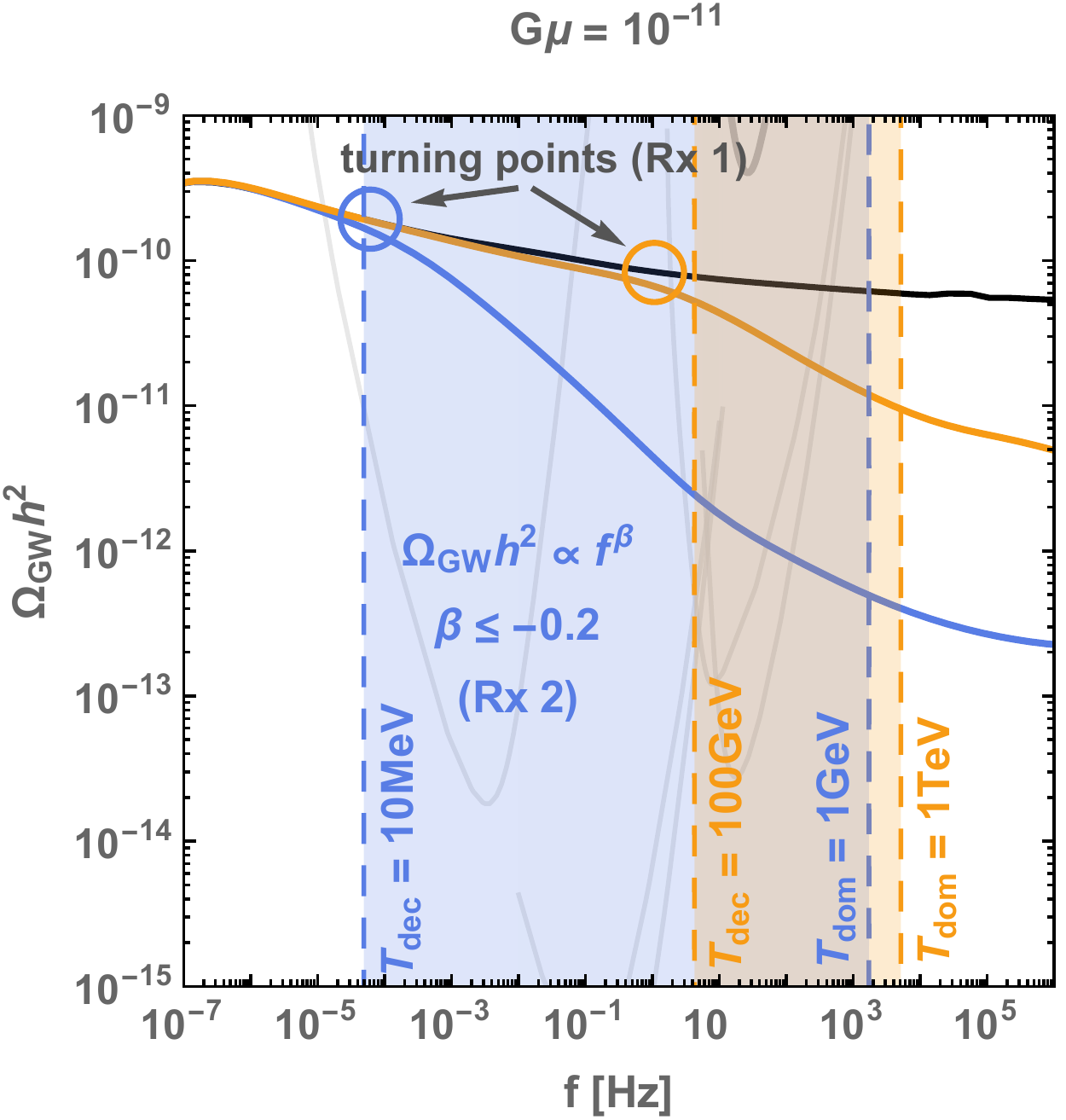}}}
\raisebox{0cm}{\makebox{\includegraphics[height=0.52\textwidth, scale=1]{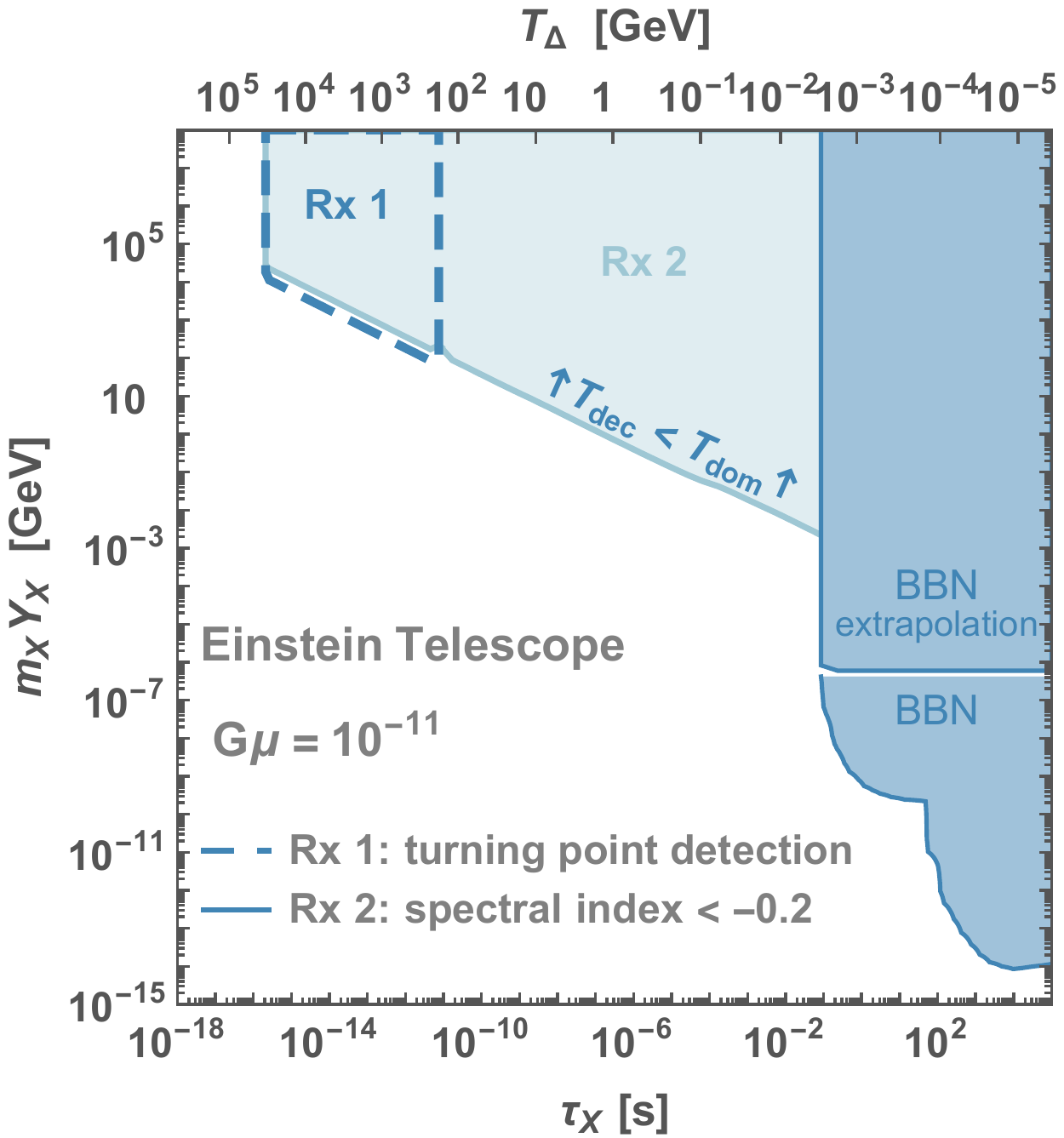}}}
\caption{\it \small \label{fig:FullMD}  \textbf{Left}: SGWB for $G\mu =10^{-11}$ assuming that a heavy cold particle dominates the energy density of the universe at the temperature $T_{\rm dom}$ and decays at the temperature $T_{\Delta} = T_{\rm dec}$. \textbf{Right}:  Considering the particular case of the Einstein Telescope, we illustrate how the constraints on the abundance and lifetime of a heavy relic depend on the choice of the prescription, Rx 1 or Rx 2 defined in Sec.~\ref{sec:triggerMatterGW}.}
\end{figure}

\subsection{How to detect a matter era with a GW interferometer}
\label{sec:triggerMatterGW}

For a first qualitative analysis, we start with two simple prescriptions for detecting a matter era from the measurement of a SGWB from CS by a GW interferometer.
\begin{itemize}
\item
\textbf{Rx 1} \textit{(turning-point prescription)}: The turning point, namely the frequency at which the spectral index of the GW spectrum changes, corresponding to the transition from the matter to the radiation era, defined in Eq.~\eqref{eq:f_delta_RD}, must be inside the interferometer window, as shown for instance  in Fig.~\ref{fig:OmegaGWCS}.
\item
\textbf{Rx 2} \textit{(spectra-index prescription)}: The measured spectral index must be smaller than $-0.2$, namely $\beta < -0.2$ where $\Omega_{\rm GW}h^2 \propto f^{\beta}$.
\end{itemize}
In Fig.~\ref{fig:FullMD}, we compare the above two prescriptions. The prescription Rx 1 is more conservative but enough to measure the lifetime of the particle. In our study, we use the prescription Rx 1 and, in Fig.~\ref{fig:sketch}, we show how to extend the constraints with Rx 2.

We note here that the presence of the turning point and the changed spectral index at high frequencies would be similar in the case of a long intermediate inflation era instead of an intermediate matter era. Disentangling the two effects  deserves further studies. Interestingly, high-frequency burst signals due to cusp formation could be a way-out  \cite{CS_VOS}. In the analysis of this paper, we interpret the suppression of the GW spectrum as due to an intermediate matter era.

\begin{figure}[p!tb]
\thispagestyle{empty}
\centering\vspace{-2.2cm}
\raisebox{0cm}{\makebox{\includegraphics[height=0.45\textwidth, scale=1]{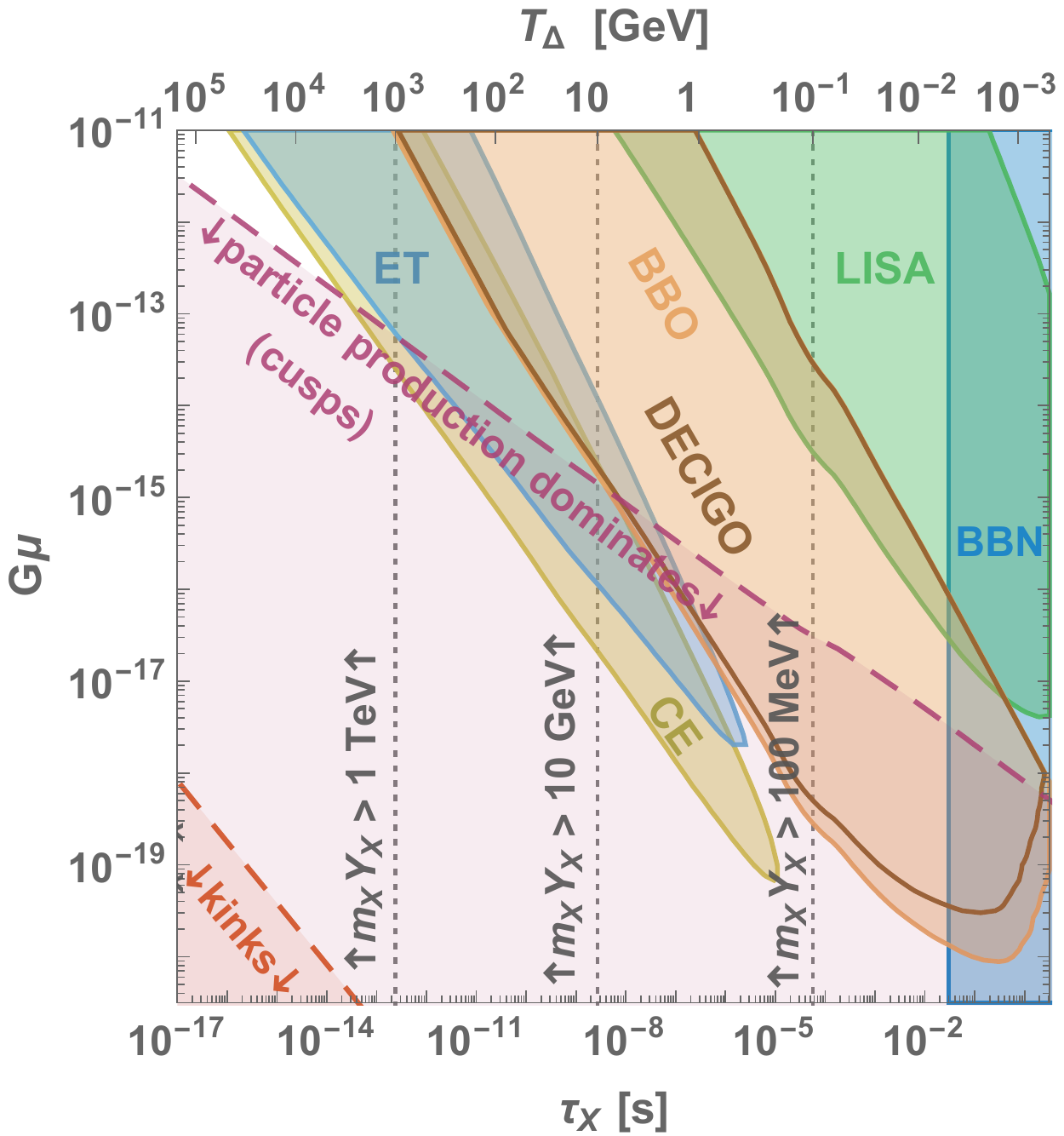}}}
\raisebox{0cm}{\makebox{\includegraphics[height=0.45\textwidth, scale=1]{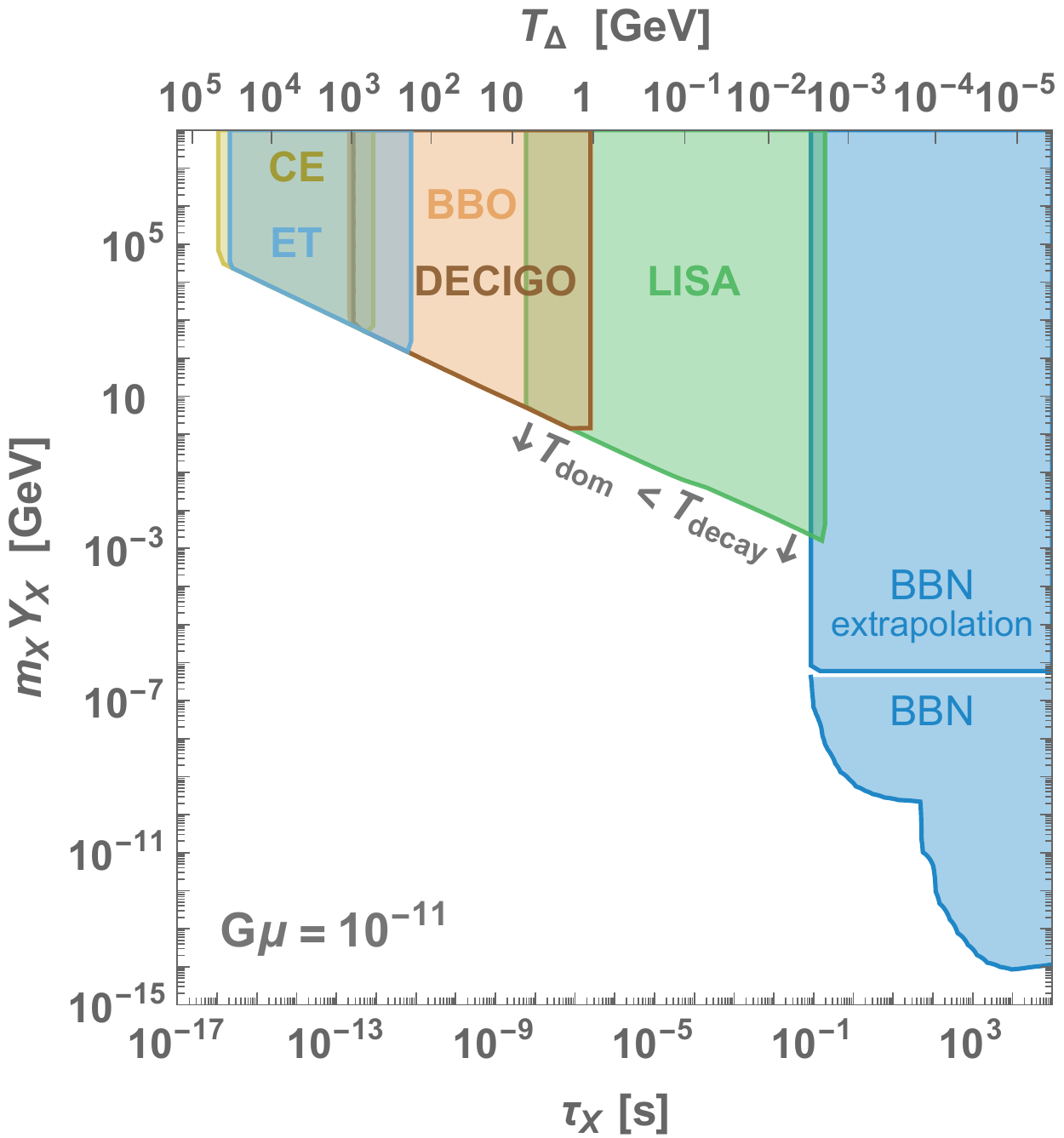}}}\vspace{0.1cm}
\raisebox{0cm}{\hspace{-0.0cm}\makebox{\includegraphics[height=0.45\textwidth, scale=1]{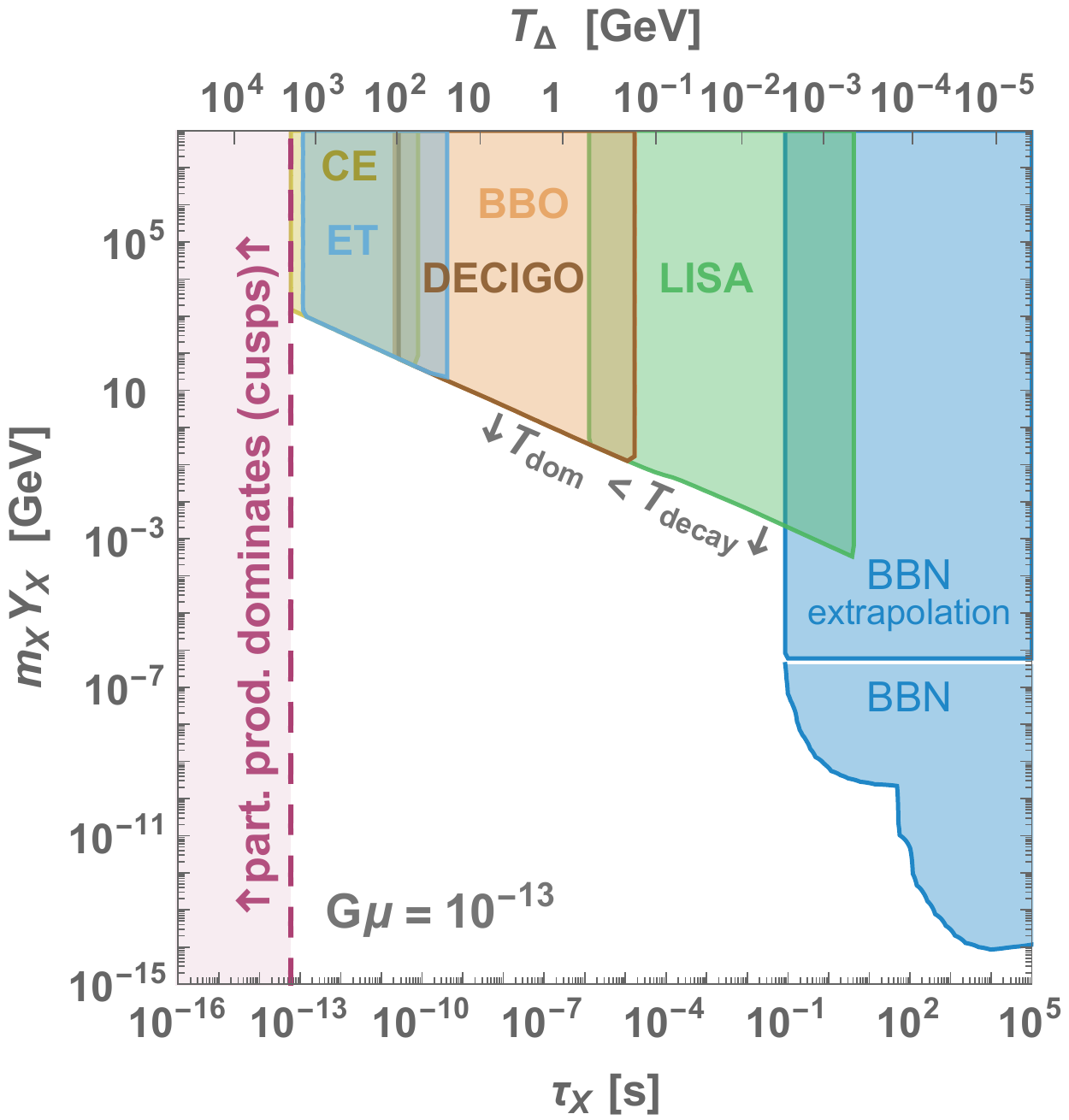}}}
\raisebox{0cm}{\hspace{0.0cm}\makebox{\includegraphics[height=0.45\textwidth, scale=1]{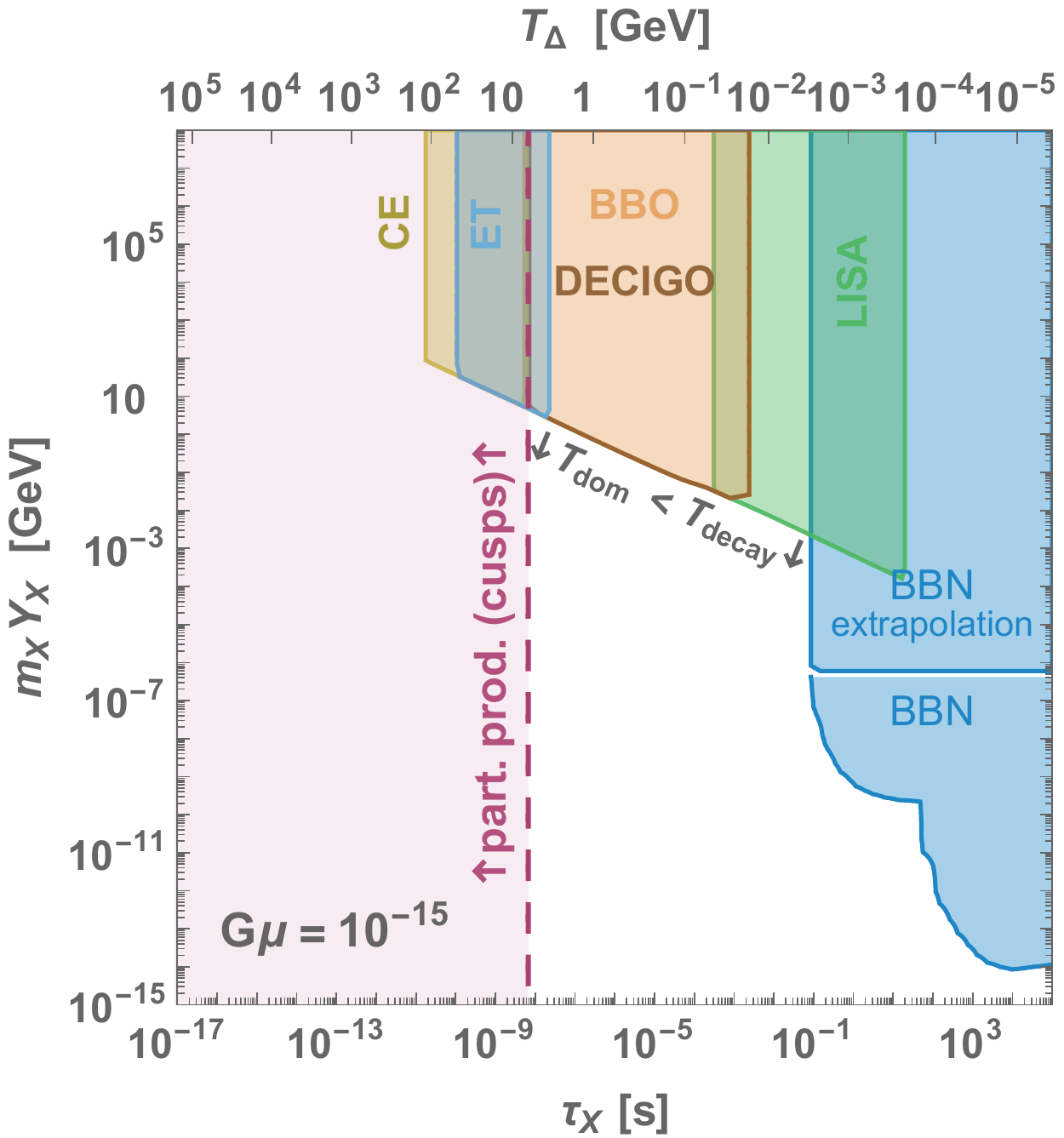}}}\vspace{0.1cm}
\raisebox{0cm}{\hspace{-0.0cm}\makebox{\includegraphics[height=0.45\textwidth, scale=1]{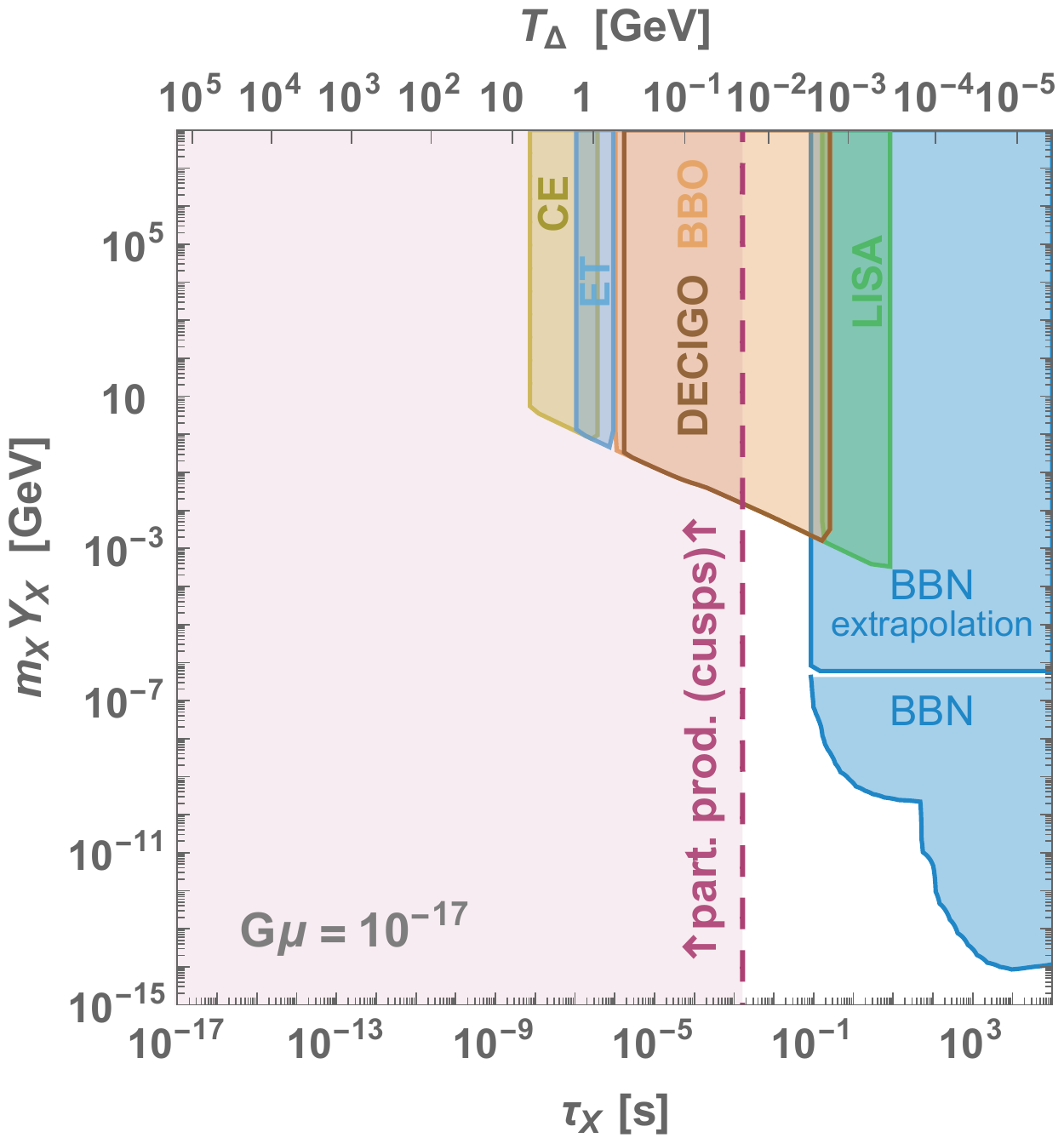}}}
\raisebox{0cm}{\hspace{0.0cm}\makebox{\includegraphics[height=0.45\textwidth, scale=1]{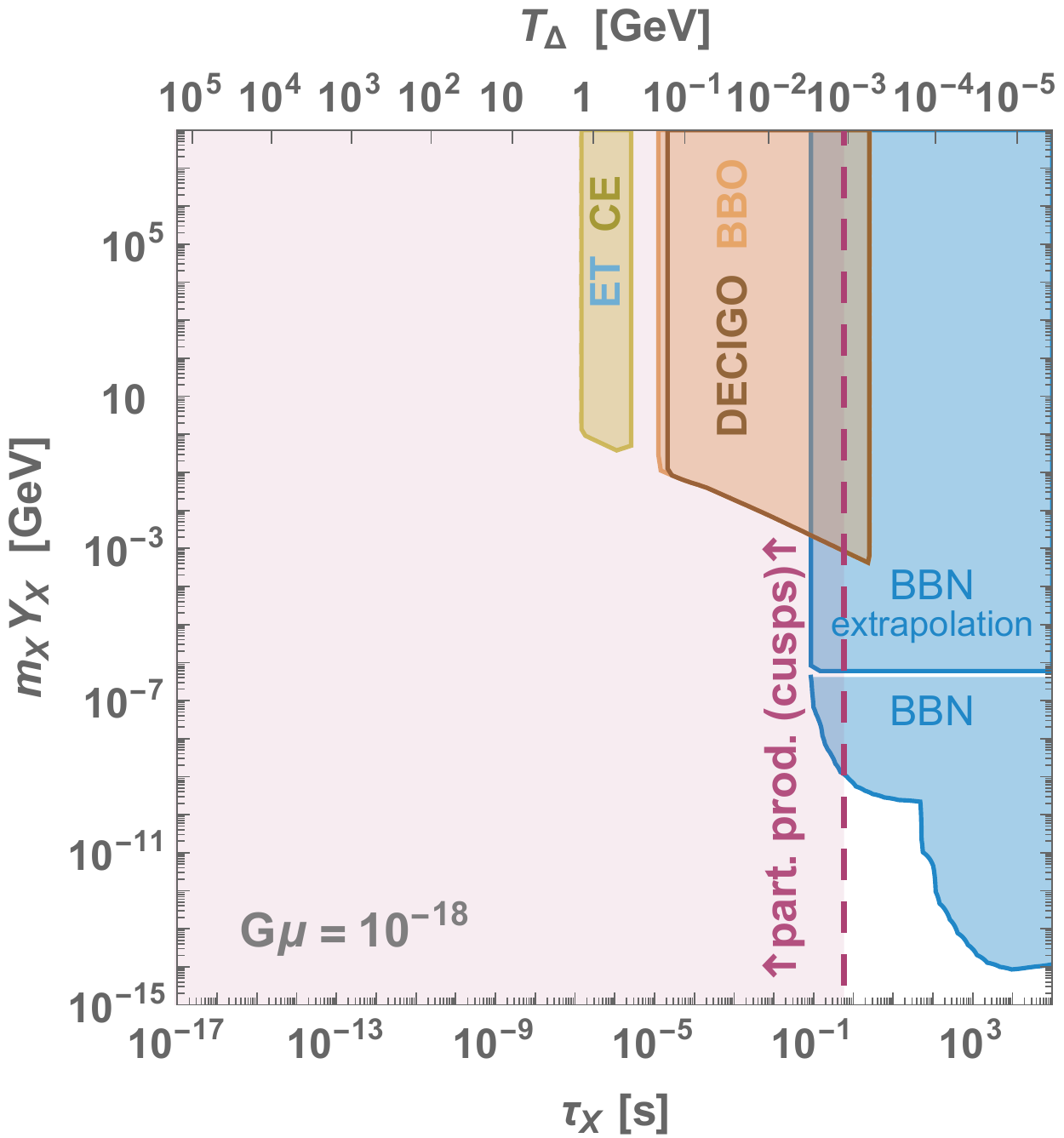}}}
\vspace{-0.3cm}
\caption{\it \small \label{fig:mY_tauX_GWI_VS_BBN} Constraints on the lifetime $\tau_X$ and would-be abundance $m_{X}Y_{X}$ of a heavy unstable particle inducing an early-matter era, assuming the observation of a SGWB from CS by a GW interferometer, c.f. Sec.~\ref{eq:model_indpt}. We compare the new prospects with the current limits inferred from BBN \cite{Jedamzik:2006xz, Jedamzik:2009uy, Kawasaki:2017bqm}. We assume the detectability of the turning point in the GW spectrum at the frequency $f_{\Delta}$, induced by the decay of the particle at $T_{\rm dec}=T_{\Delta}$, c.f. \textit{turning-point description Rx 1} in Sec.~\ref{sec:triggerMatterGW}. Limitation due to particle production in the cusp-domination case \cite{CS_VOS} are shown in purple.}
\end{figure}

\subsection{Model-independent constraints on particle physics parameters}
\label{eq:model_indpt}

A matter-dominated era may result from an oscillating scalar field  \cite{Turner:1983he}, such as a moduli field, or a relativistic plasma with a non-vanishing tensor bulk viscosity \cite{ Boyle:2007zx}, or simply a massive particle dominating the energy density of the universe. A matter-dominated era may be motivated by the possibillity to enhance structure growth at small scales, since density perturbations start to grow linearly earlier \cite{Erickcek:2011us, Fan:2014zua}, hence boosting the dark matter indirect detection signals  \cite{Delos:2019dyh,Blanco:2019eij}, or the possibility to enhance the primordial black holes production \cite{Polnarev:1986bi, Green:1997pr, Georg:2016yxa}.

\begin{figure}[h!]
\centering
\raisebox{0cm}{\makebox{\includegraphics[height=0.6\textwidth, scale=1]{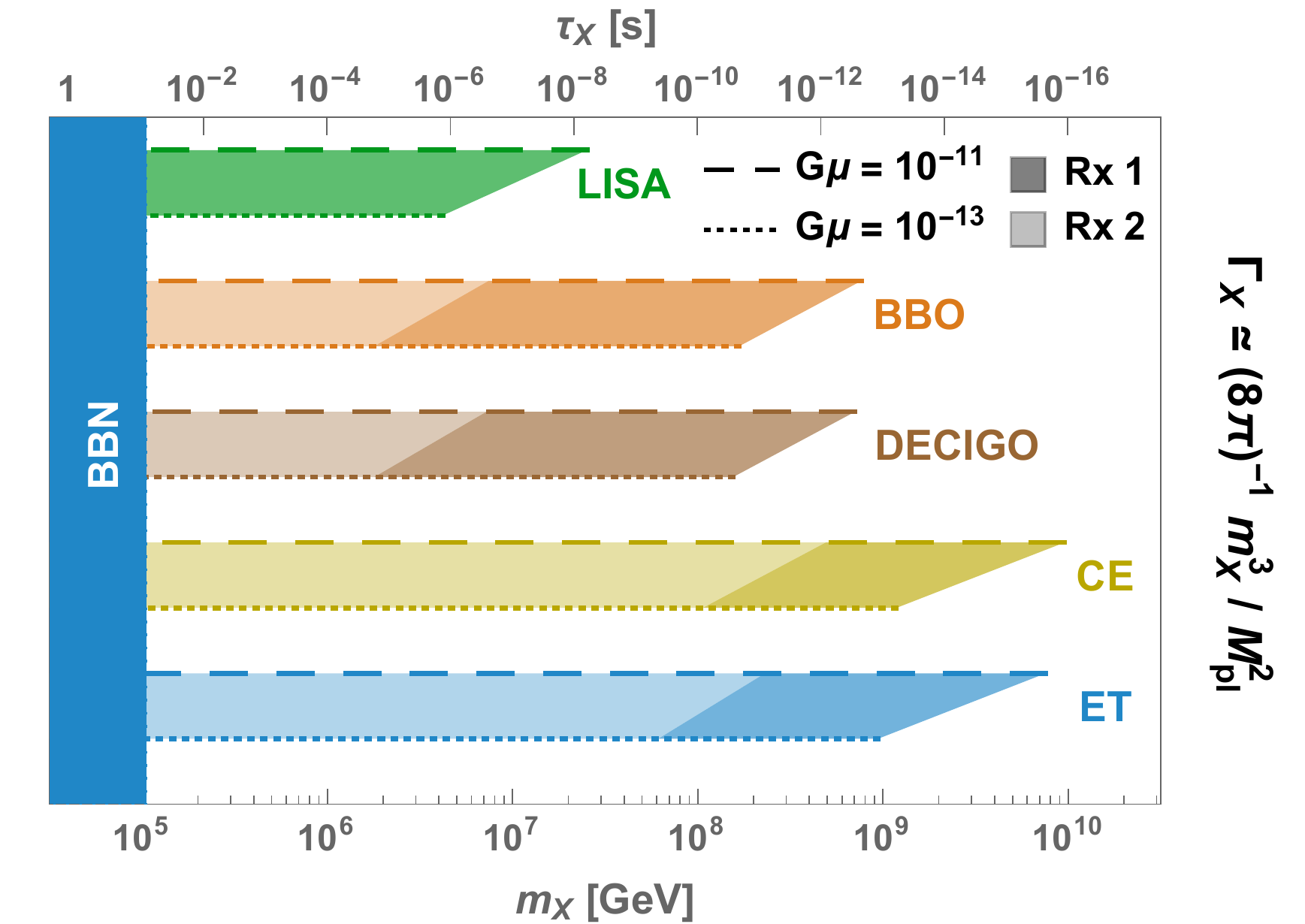}}}
\caption{\it \small \label{fig:sketch}Reach of future GW interferometers on the mass of a  heavy particle decaying through a Planck suppressed operator, $\Gamma_{X} \propto m_{X}^3/M_{\rm pl}^2$, supposing that it is sufficiently produced to induce a matter era before the decay. We compare the \textit{turning-point prescription (Rx 1)} and the \textit{spectral-index prescription (Rx 2)} discussed in Sec.~\ref{sec:triggerMatterGW}. { In Sec.~\ref{sec:BenchmarkModels}, we study three different production mechanisms of such particle with Planck suppressed decay width: scalar oscillating moduli produced after supersymmetry breaking in Sec.~\ref{sec:moduli}, scalar particle gravitationally produced at the end of inflation in Sec.~\ref{sec:ScalarGrav}, or scalar particle produced via thermal freeze-in assuming a Higgs-mixing in Sec.~\ref{sec:ScalarHiggs}.} }
\end{figure}

We suppose an early-matter era is caused by the energy density of a cold particle $X$, meaning that $X$ is non-relativistic and is decoupled chemically and kinetically from the visible sector. The energy density $m_X n_X$ of $X$ dominates over the energy density of the SM radiation, with entropy $s_{\rm SM}$, at the temperature $T_{\rm dom}$
\begin{equation}
\label{eq:Tdom_def}
T_{\rm dom}=\frac{4}{3} m_{X} Y_{X}, \qquad Y_{X} \equiv n_{X}/s_{\rm SM}.
\end{equation}
Then, the cold relic decays when  its lifetime $\tau_X$ is equal to the age of the universe, corresponding to the temperature
\begin{equation}
\label{eq:T_dec_tauX}
T_{\rm dec} =1~\text{GeV} \left( \frac{80}{g_{\rm SM}}\right)^{1/4}\left( \frac{2.7\times 10^{-7}~\text{s}}{\tau_X} \right)^{1/2}.
\end{equation}
Note that the above relation between $T_{\rm dec}$ and $\tau_X$ only assumes that the decay is followed by a radiation dominated era and is independent of the previous thermal history of the universe.
$T_{\rm dec}$ is sometimes referred, mistakenly though \cite{Scherrer:1984fd}, as the reheating temperature following the decay.
We propose to use the third generation of GW interferometers to constrain cold relics responsible for early-matter domination. 
The constraints we will derive rely on the following assumptions:
\begin{itemize}
\item[1)] A SGWB from CS with tension $G\mu$ is measured by a GW interferometer $i$. 

\item[2)] The cold particle is abundant enough to lead to a matter-dominated era before it decays
\begin{equation}
T_{\rm dom} > T_{\rm dec} \ ,
\end{equation} 
where $T_{\rm dom}$ and $T_{\rm dec}$ satisfy Eq.~\eqref{eq:Tdom_def} and Eq.~\eqref{eq:T_dec_tauX}.

\item[3)]
The prescription Rx 1 of Sec.~\ref{sec:triggerMatterGW} is used, i.e.  the turning point in the GW spectrum  is in the observation window of the detector and 
\begin{equation}
\Omega_{GW}(f_{\Delta}(T_{\rm dec}, \, G\mu), \, G\mu) h^2 \, > \,  \Omega_{\rm sens}^{(i)}h^2,  
\end{equation}
where $\Omega_{GW}(f,\, G\mu)h^2$ is the predicted scale-invariant GW spectrum  from  Eq.~\eqref{eq:spect_master}, and $ \Omega_{\rm sens}^{(i)}h^2$ is the power-law sensitivity curve of the detector $i$. 

\end{itemize}
Fig.~\ref{fig:mY_tauX_GWI_VS_BBN} shows  these new constraints  in comparison with the  current complementary constraints from BBN, usually represented in the plane $(\tau_{X}, \, m_{X}Y_{X})$ \cite{Jedamzik:2006xz, Jedamzik:2009uy, Kawasaki:2017bqm}.
We can translate the sensitivity of each interferometer to probe the particle lifetime into typical mass windows, assuming some decay width. This is illustrated 
in Fig.~\ref{fig:sketch} with a Planck-suppressed decay width $\Gamma_{X} \propto m_{X}^3/M_{\rm pl}^2$.
%
%
%

\section{Application to benchmark models}
\label{sec:BenchmarkModels}

\subsection{Oscillating scalar moduli}
\label{sec:moduli}

\begin{figure}[h!]
\centering
\raisebox{0cm}{\makebox{\includegraphics[height=0.51\textwidth, scale=1]{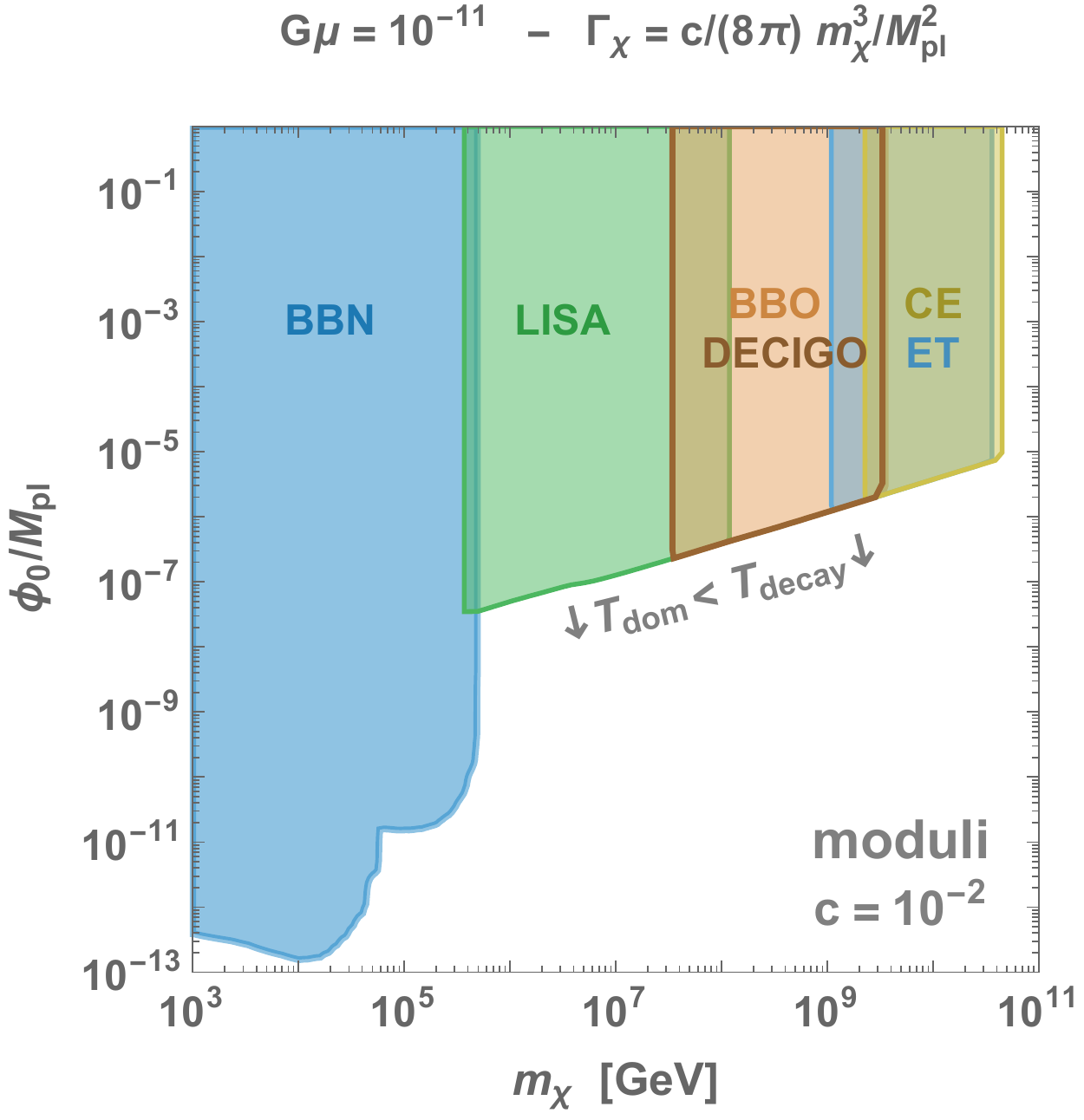}}}
\raisebox{0cm}{\makebox{\includegraphics[height=0.51\textwidth, scale=1]{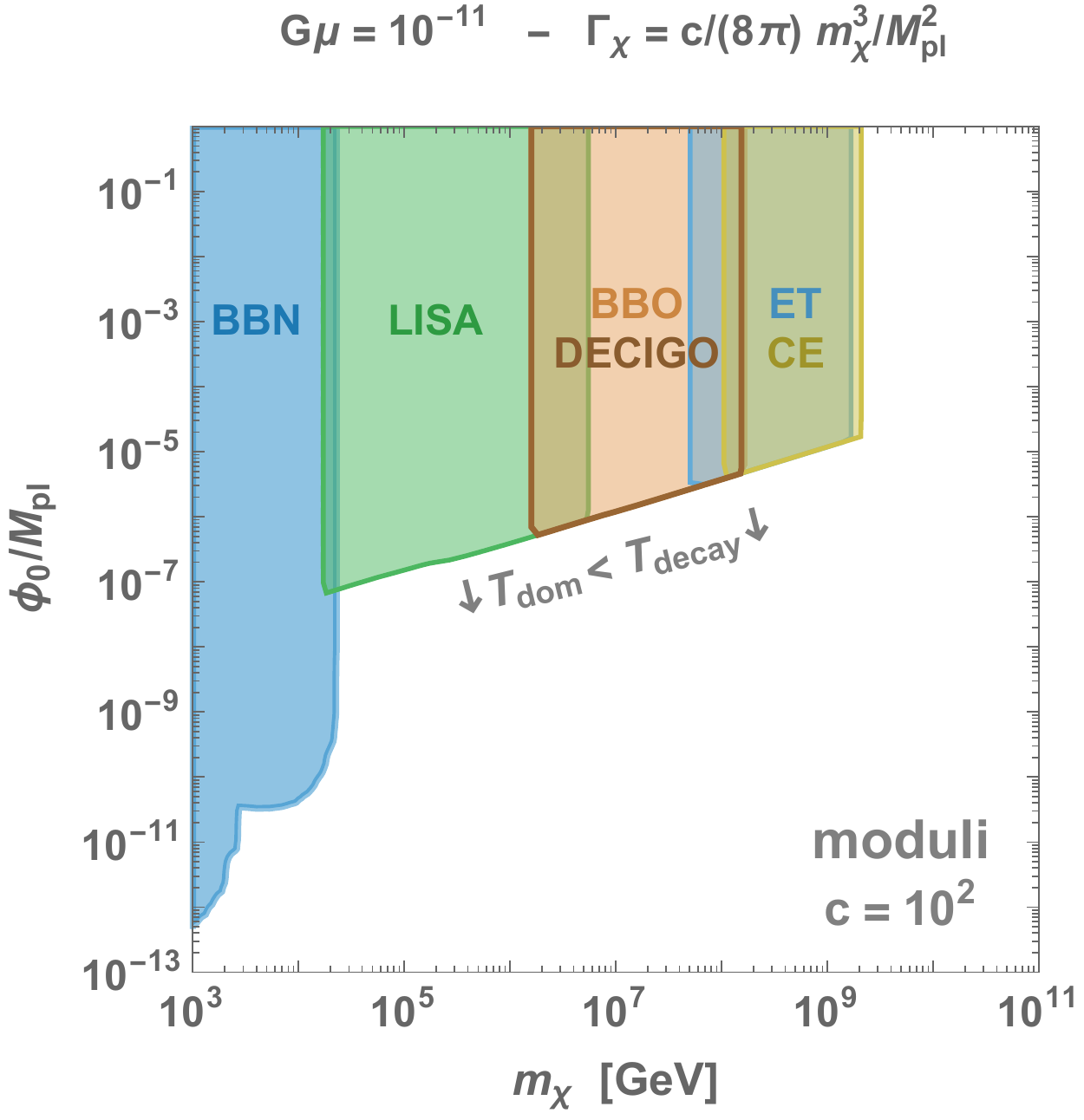}}}
\caption{\it \small \label{fig:GWconst_moduli}Constraints on a moduli field oscillating in the early universe around its minima with initial amplitude $\phi_0$ and mass $m_{\phi}$, from the non-observation of its signature in the GW spectrum from CS with tension $G\mu = 10^{-11}$. In stringy UV completions, the lightest moduli field is related to the gravitino scale $m_{X} \sim m_{3/2}$ \cite{Acharya:2010af}. The BBN constraints are taken from \cite{Jedamzik:2006xz, Jedamzik:2009uy, Kawasaki:2017bqm}. We use the \textit{turning-point prescription (Rx 1)} discussed in Sec.~\ref{sec:triggerMatterGW}. Constraints are tightened when considering the \textit{spectral-index prescription (Rx 2)}, as illustrated in Fig.~\ref{fig:sketch}.}
\end{figure}

String theory vacua feature moduli fields which characterize the size and shape of the compactification manifold. From a 4D effective field theory perspective, they are fields with flat potential e.g. axions or dilatons. After supersymmetry (SUSY) breaking, one expects moduli fields to acquire a mass of the order of the gravitino mass scale for the lightest \cite{Acharya:2010af}, e.g. $m_{3/2} \sim $~TeV for low-scale SUSY. As soon as the Hubble rate satisfies $H \lesssim m_{\phi}$, the scalar field starts coherent oscillations and its energy density redshifts as matter. We assume that the onset of oscillations  occurs during radiation domination, at the temperature $T_{\rm osc}$ 
\begin{equation}
\frac{\pi^2g_{*}T_{\rm osc}^4}{90\,M_{\rm pl}} \equiv m_{\phi}^2
\end{equation}
where we fix the number of relativistic degrees of freedom to the fiducial value $g_{*}=106.75$.
Then, the moduli starts dominating the energy density of universe at the temperature, c.f. Eq.~\eqref{eq:Tdom_def}
\begin{equation}
T_{\rm dom} = \frac{4}{3} \frac{\rho_{\phi}^{\rm osc}}{s_{\rm osc}} \equiv \frac{\tfrac{1}{2} m_{\phi}^2\phi_{0}^2}{\tfrac{2\pi^2}{45}g_{*}T_{\rm osc}^3},
\end{equation}
where $\phi_{0}$ is the vacuum expectation value of the moduli field when it starts to oscillate.
For concreteness, we consider moduli fields which interact with the visible sector via 
Planck-suppressed operators and hence have decay widths of order
\begin{equation}
\label{eq:moduli_decay_width}
\Gamma_{\phi} \simeq \frac{c}{8\pi}\frac{m_{\phi}^3}{M_{\rm pl}^2},
\end{equation}
where  $M_{\rm pl} \simeq 2.4 \times 10^{-18}$~GeV and $c$ is a model-dependent factor which we suppose to be in the range $10^{-2} \lesssim c \lesssim 10^2$. For TeV-scale moduli mass, the moduli lifetime is long, $\Gamma_{\phi}^{-1}\sim 10^5$~s, and the decay occurs much after BBN. Imposing that the energy density of the moduli decay products, $\rho_{\phi}$, is smaller than a fraction $10^{-14}$ of the total entropy density of the universe in order to preserve the predictions of BBN, c.f. Fig.~\ref{fig:mY_tauX_GWI_VS_BBN} and \cite{Jedamzik:2006xz, Jedamzik:2009uy, Kawasaki:2017bqm}, one constrains the vacuum expectation value of the moduli field, just after it starts oscillating, to be, c.f. Fig.~\ref{fig:GWconst_moduli}
\begin{equation}
\label{eq:BBN_moduli}
\text{BBN is preserved for TeV-scale moduli:} \qquad \phi_{0} \lesssim 10^{-12} ~ M_{\rm pl}.
\end{equation}
A large moduli VEV is expected from the dependence of the moduli potential on the inflaton VEV \cite{Dine:1983ys, Coughlan:1984yk, Dine:1995kz, Acharya:2008bk}, except if the scalar moduli field lies at a point of enhanced symmetry where the induced minimum at late times coincides with the minimum at earlier times \cite{Dine:1995uk}.
However, even in the case where the moduli VEV after SUSY breaking remains small, one expects moduli to be copiously produced both through thermal \cite{Ellis:1982yb, Nanopoulos:1983up, Ellis:1984eq} and gravitational production, c.f. Sec.~\ref{sec:ScalarGrav} and \cite{Giudice:1999yt,Felder:1999wt}, hence violating the bound in Eq.~\eqref{eq:BBN_moduli}.

%

The scalar moduli problem is similar to the gravitino problem \cite{Weinberg:1982zq}, both are copiously produced relics, with weak-scale mass and Planck-suppressed decay rate,  spoiling the BBN predictions. An important difference is that during inflation, the energy density is diluted in the fermionic case but is frozen in the scalar case as long as $H \gg m_{\phi}$. Hence, as opposed to the gravitino case, only a low-scale (weak-scale) inflation can exponentially dilute the scalar field and solve the moduli problem \cite{Randall:1994fr}. Other proposed solutions are to increase the moduli mass up to $m_{\phi}\sim O(10^3)\,m_{3/2}$, c.f. Fig.~\ref{fig:GWconst_moduli} and \cite{Binetruy:1997vr}, or to form substructures (modular stars) which enhances the decay \cite{Banks:1995dt, Krippendorf:2018tei}, or to produce gauge fields from the tachyonic instability \cite{Giblin:2017wlo}.

A pragmatic approach to solve the moduli problem is to break SUSY at a much higher scale, at the expense of large fine-tuning, like in so-called  High-scale SUSY \cite{Hall:2009nd} or Split SUSY \cite{Wells:2004di}. 
A larger SUSY breaking scale improves gauge coupling unification \cite{ArkaniHamed:2004fb}, is compatible with the Higgs at $125$~GeV \cite{Bagnaschi:2014rsa} and is free from the main difficulties encountered by low-scale SUSY such as large flavor and CP violation \cite{ArkaniHamed:2004yi}. 
In Split  and High-scale SUSY, a $125$~GeV Higgs is compatible  with a SUSY breaking scale as high as $10^8$~GeV and  $10^{12}$~GeV respectively \cite{Bagnaschi:2014rsa}. Moduli fields with masses of the same order would then induce an early matter era which could lead to detectable features in the GW background.

As shown in Fig.~\ref{fig:GWconst_moduli}, the observation of a SGWB from CS with one of the next generation GW interferometers would provide constraints on moduli masses up to $\lesssim 10^{10}$~GeV, well above the current $\lesssim 100$~TeV currently probed by BBN. Hence, if detected, GW from CS would be a promising tool to probe superstring theories.

In addition to being naturally motivated in SUSY constructions, moduli fields have interesting cosmological consequences: Afflect-Dine baryogenesis \cite{Affleck:1984fy, Dine:1995kz},  non-thermal production of Wino-DM  \cite{Moroi:1999zb, Acharya:2008bk,Fan:2013faa}, formation of oscillons or Q-balls \cite{Krippendorf:2018tei},  the required entropy injection to allow thermal DM much heavier than the standard unitarity bound $\sim 100$~TeV \cite{Cirelli:2018iax} or to revive Grand-Unified-Theory-scale QCD axion DM \cite{Acharya:2010zx}, see Ref.~\cite{Kane:2015jia} for a review on the moduli problem and its cosmological implications.

\subsection{Scalar particles produced gravitationally}
\label{sec:ScalarGrav}

{
In the previous subsection, Sec.~\ref{sec:moduli}, we considered a model of gravitationally-only interacting particle whose abundance is given by the misalignement mechanism after SUSY breaking. Instead, we now consider the possibility to produce such a particle, gravitationnally only, at the end of inflation.
In the next subsection, Sec.~\ref{sec:ScalarHiggs}, we also consider the possibility of a thermal production via freeze-in through a Higgs mixing in the case of the conformal scalar $\xi = 1/6$ where the gravitational production is too small to lead to an early matter domination. 
}

A massive particle can be gravitationally produced at the end of the inflation due to the non-adiabatic change of its curvature-induced mass from deep de Sitter to deep radiation-domination \cite{Parker:1969au,Zeldovich:1971mw, Birrell:1982ix, Ford:1986sy,Kofman:1997yn}, hence possibly leading to heavy dark matter WIMPzillas \cite{Chung:1998zb,Kolb:1998ki, Chung:2001cb, Kolb:2017jvz, Ema:2018ucl}. Our interest here is not to explain DM but to predict a non-standard matter era in the early universe. If coupled non-conformally to gravity and if the condition of non-adiabaticity, $m_X \lesssim H_{\rm inf}$, is satisfied, then the particle $\chi$ will be produced abundantly,  potentially leading to an early matter domination. The lagrangian is
\begin{equation}
\label{eq:action_wimpzilla}
\mathcal{S} = \int d^4x \sqrt{-g} \left( \frac{1}{2}\left(M_{\mathsmaller{\rm P}}^2  - \xi \chi^2\right)R - \frac{1}{2}g^{\mu\nu}\partial_{\mu}\chi\partial_{\nu}\chi - \frac{1}{2}m_{\chi}^2\chi^2 \right],
\end{equation}
with $\xi$ the non-minimal coupling to gravity. We consider the cases $\xi =0$ (minimal coupling) and $\xi =1/6$ (conformal coupling).
We suppose that the scalar $\chi$ decays gravitationally through Planck suppressed operators
\begin{equation}
\label{eq:decay_width_grav}
\Gamma_{\chi} \simeq \frac{1}{8\pi}\frac{m_{\chi}^3}{M_{\rm P}^2}.
\end{equation}
A too light scalar would spoil the BBN prediction.
The comoving number density, $Y_{\chi}\equiv n_{\chi}/s$,  of a minimal scalar, $\xi = 0$, after gravitational production is \cite{Kolb:2017jvz}
\begin{equation}
Y_{\chi}^{\xi =0} \simeq \dfrac{H_{\rm reh}^2 H_{\rm inf}}{s} \left\{
                \begin{array}{ll}
                 96 \; \dfrac{H_{\rm inf}}{m_{\chi}} \hspace{2.75 cm} \dfrac{m_{\chi}}{H_{\rm inf}} < 1, \vspace{0.25 cm} \\ 
                0.76 \; \dfrac{H_{\rm inf}}{m_{\chi}} e^{-2 m_{\chi}/H_{\rm inf}} \qquad \dfrac{m_{\chi}}{H_{\rm inf}} > 1,
                \end{array}
              \right.
\end{equation}
\begin{figure}[b!]
\centering
\vspace{0cm}
\raisebox{0cm}{\makebox{\includegraphics[height=0.6\textwidth, scale=1]{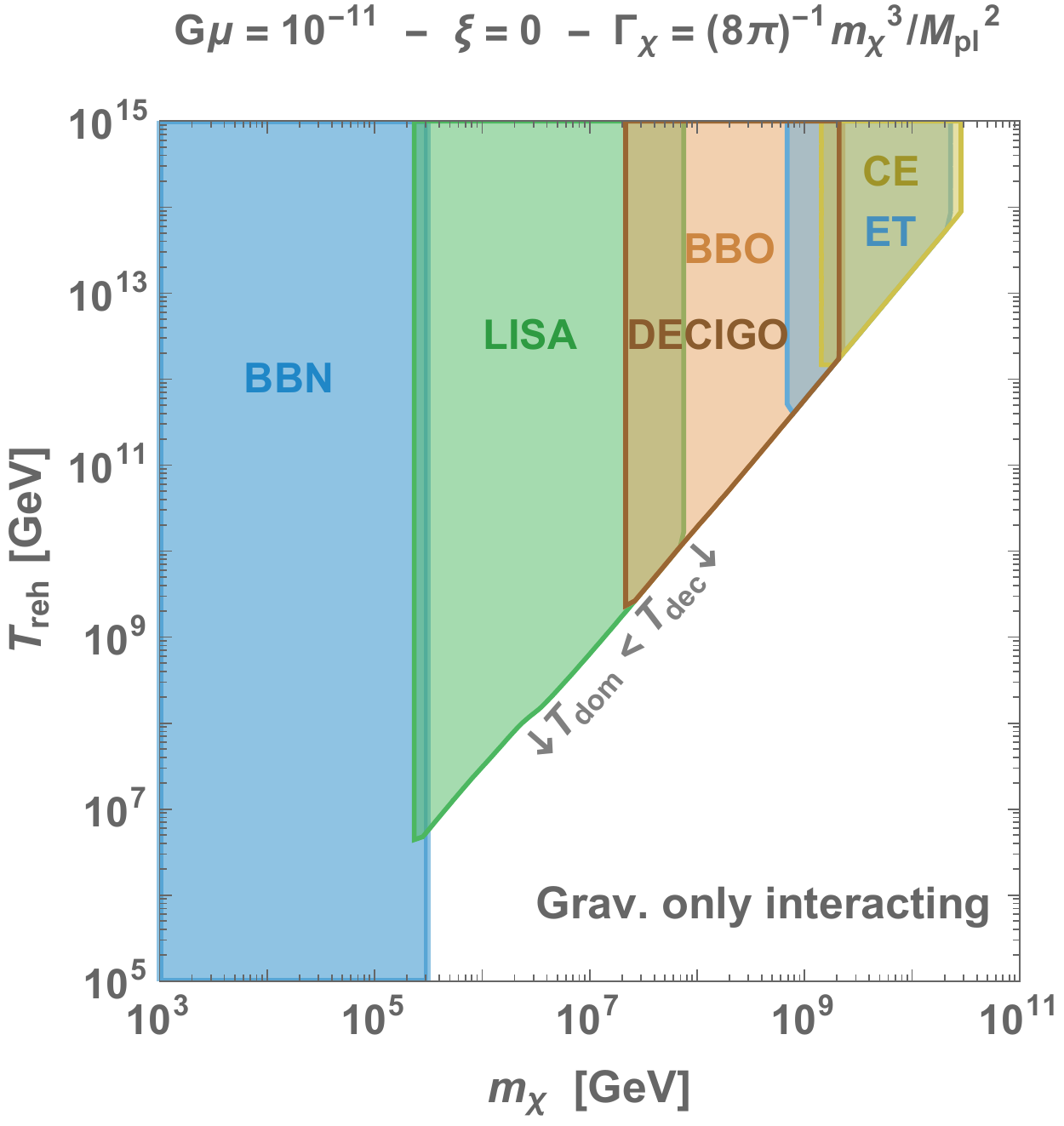}}}
\caption{\it \small \label{fig:GWconst_OnlyGravCoupled} Constraints on a purely gravitationally produced ($\kappa = 0$) non-conformal scalar ($\xi = 0$) decaying via a Planck suppressed operator, c.f. Sec.~\ref{sec:ScalarGrav}, assuming the observation of a SGWB from CS with tension $G\mu = 10^{-11}$ by third-generation GW detectors.  We have fixed the inflation scale $H_{\rm inf} = 10^{13}$~GeV. We use the \textit{turning-point prescription (Rx 1)} discussed in Sec.~\ref{sec:triggerMatterGW}.   Constraints are tightened when considering the \textit{spectral-index prescription (Rx 2)}, as shown in Fig.~\ref{fig:sketch}.}
\end{figure}
\begin{figure}[h!]
\centering
\vspace{0cm}
\raisebox{0cm}{\makebox{\includegraphics[height=0.6\textwidth, scale=1]{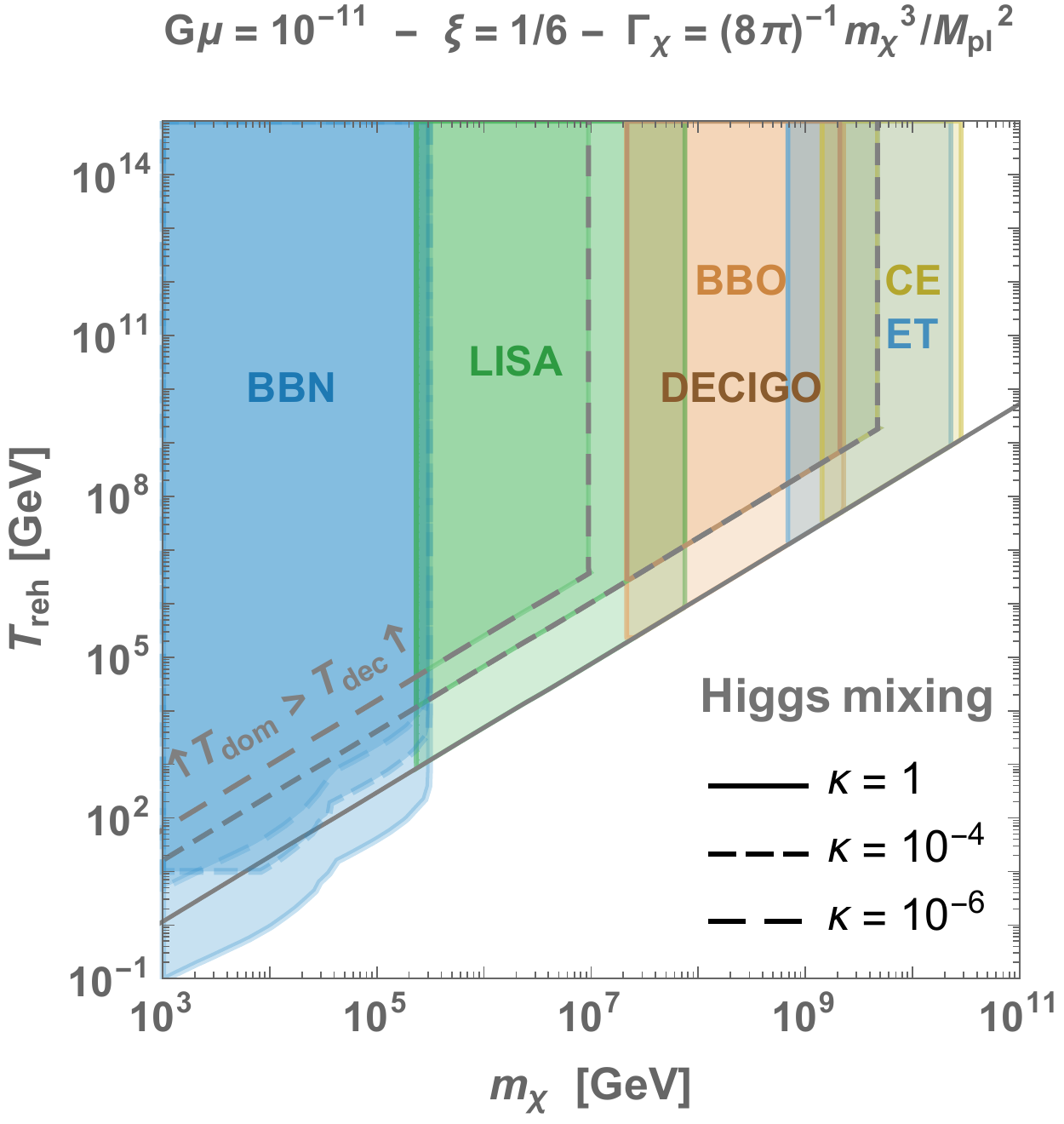}}}
\caption{\it \small \label{fig:GWconst_OnlyGravCoupled2} Constraints on a scalar conformally coupled to gravity, $\xi = 1/6$, thermally produced via a mixing with the Higgs $\kappa = 10^{-6},10^{-4},1$, c.f. Sec.~\ref{sec:ScalarHiggs}, from the measurement of a SGWB from CS. We have also included the gravitational production, c.f. \ref{sec:ScalarGrav}, even though it is negligible. 
We have fixed the inflation scale $H_{\rm inf} = 10^{13}$~GeV. We use the \textit{turning-point prescription (Rx 1)} discussed in Sec.~\ref{sec:triggerMatterGW}. Constraints are tightened when considering the \textit{spectral-index prescription (Rx 2)}, as shown in Fig.~\ref{fig:sketch}.}
\end{figure}
where $H_{\rm inf}$ and $H_{\rm reh}$ are the Hubble factors at the end of inflation and reheating. We have checked that particle production caused by the oscillations of the inflaton during preheating, potentially relevant when $H_{\rm inf} \lesssim m_{\chi} \lesssim m_{\phi}$ where $m_{\phi}$ is the inflaton mass \cite{Ema:2015dka, Ema:2016hlw, Ema:2018ucl, Ema:2019yrd, Chung:2018ayg}, is not strong enough to ignite an early matter domination before BBN.

For our study, we fix the inflation scale $H_{\rm inf} = 10^{13}$~GeV close to its upper bound value $6 \times 10^{13}$~GeV from the non-detection of the fundamental B-mode polarization patterns in the CMB \cite{Ade:2018gkx, Akrami:2018odb}. 
If produced with a sufficient amount, the scalar field can lead to an early-matter domination era. In Fig.~\ref{fig:GWconst_OnlyGravCoupled}, we show the GW constraints on the $\chi$ scalar particle, which is non-conformally coupled to gravity. We see that for reheating temperature larger than $10^7$~GeV, the scalar field is sufficiently produced by gravitational effects at the end of inflation to dominate the energy density of the universe before BBN starts, and be detected in GW experiments. Masses as large as $10^{10}$~GeV can be probed.

The gravitational production of particles conformally coupled to gravity, i.e. scalars with $\xi=1/6$, transverse vectors or fermions, is less efficient than for a minimal scalar, $\xi = 0$ \cite{Dimopoulos:2006ms, Chung:2011ck, Graham:2015rva, Kolb:2017jvz, Ema:2019yrd}. Here we give the abundance computed in \cite{Kolb:2017jvz} for a conformal scalar\footnote{Here `conformal scalar' means `conformally coupled to gravity' with $\xi = 1/6$. In any case, the conformal symmetry is broken via the scalar mass term.}
\begin{equation}
Y_{\chi}^{\xi = 1/6} \simeq \dfrac{H_{\rm reh}^2 H_{\rm inf}}{s} \left\{
                \begin{array}{ll}
                 0.0010 \; \dfrac{m_{\chi}}{H_{\rm inf}} \hspace{2.5 cm} \dfrac{m_{\chi}}{H_{\rm inf}} < 1, \vspace{0.25 cm} \\ 
                0.0040 \; \dfrac{H_{\rm inf}}{m_{\chi}} e^{-2 m_{\chi}/H_{\rm inf}} \qquad \dfrac{m_{\chi}}{H_{\rm inf}} > 1,
                \end{array}
              \right.
\end{equation}
We check that the gravitational production of such particles, conformally-coupled to gravity, is not strong enough to lead to a matter-domination era before BBN starts, if the reheating temperature following inflation is below $10^{13}~$GeV. Hence, in the next section we consider another production mechanism by introducing a mixing with the standard model Higgs.

\subsection{Scalar particles produced through the Higgs portal }
\label{sec:ScalarHiggs}

The gravitational production of a scalar conformally-coupled to gravity ($\xi = 1/6$), is too small to lead to an early-matter domination. 
On the other hand, a mixing with the Higgs $H$
\begin{equation}
\mathcal{L} \supset \frac{\kappa}{2} \chi^2 |H|^2,
\end{equation}
can lead to a large abundance via thermal freeze-in \cite{Kolb:2017jvz}
\begin{equation}
Y_{\chi} \simeq \dfrac{H_{\rm reh}^2 H_{\rm inf}}{s} \left\{
                \begin{array}{ll}
                 \dfrac{105 |\kappa|^2}{64\pi^4} \dfrac{T_{\rm max}^{12}}{H_{\rm inf}^4\, m_{\chi}^8} e^{-2m_{\chi}/T_{\rm max}} f_{0}(m_{\chi}/T_{\rm max}) \hspace{1.4 cm} T_{\rm reh} \ll m_{\chi}, \vspace{0.25 cm} \\ 
                \dfrac{ 3|\kappa|^2}{2048\pi^3} \dfrac{T_{\rm max}^{12}}{ T_{\rm reh}^7 \,H_{\rm inf}^4 \, m_{\chi} } \hspace{5 cm}  m_{\chi} \ll T_{\rm reh},
                \end{array}
              \right.
\end{equation}
where $T_{\rm max} = T_{\rm reh} \, (H_{\rm inf} / H_{\rm reh})^{1/4}$ and $T_{\rm reh} $ are respectively the maximal temperature after inflation and the reheating temperature, and $f_0(x)\equiv 1+2x + 2x^2 + 4x^3/3 + 2x^4/3 +4x^5/15+4x^6/45 + 8x^7/315 + 2x^8/315$.
In Fig.~\ref{fig:GWconst_OnlyGravCoupled2}, we show that a reheating temperature as low as $10^3$~GeV (for $\kappa =1$) can induce a matter-domination era before BBN  and leave an imprint in the would-be SGWB from CS detectable by GW interferometers. 
For $\kappa =1$, masses as large as $10^{10}$~GeV can be probed.


\subsection{Heavy dark photons}
\label{sec:U(1)_model}

\paragraph{The $U(1)_D$ dark photon:}

We consider a $U(1)_{\mathsmaller{\rm D}}$ gauge boson, $V_{\mu}$, the dark photon, of mass $\mV$, kinematically coupled to the $U(1)_{\mathsmaller{\rm Y}}$ gauge boson of the SM \cite{Holdom:1985ag, Foot:1991kb}
\begin{equation}
\mathcal{L} \supset - \frac{\epsilon}{2c_{w}} F_{\mathsmaller{\rm Y}\mu\nu} F_{\mathsmaller{\rm D}}^{\mu\nu},
\label{lagrangian_U(1)_D_model_DP}
\end{equation}
where $c_w$ is the cosine of the weak angle and $\epsilon$ is the dark-SM coupling constant. The decay width into SM, $\Gamma_{V}$, is computed in \cite{Cirelli:2018iax}. We here report the expression for $\mV \gtrsim 2 m_{Z}$
\begin{equation}
\Gamma_{V} \simeq \left(3\times 10^{-8}~\textrm{s}\right)^{-1} \, \left(\frac{\epsilon}{10^{-9}}\right)^2 \left( \frac{\mV}{1~\text{TeV}}\right).
\end{equation}
The dark photon leads to an early-matter-dominated era if it has a large energy density $\mV\, Y_V \gtrsim 10$ GeV and a long lifetime $\tauV \sim 10^{-8}$~s, c.f. Fig.~\ref{fig:mY_tauX_GWI_VS_BBN} at $G\mu = 10^{-11}$.
Supposing that the dark photon abundance is close to thermal, $Y_{\mathsmaller{\rm V}} \sim 0.02$, c.f. Eq.~\eqref{eq:DP_comoving_number}, this implies $\epsilon \lesssim 10^{-9}$. At such a low $\epsilon$, the dark sector and the SM sector may have never been at thermal equilibrium (c.f. \cite{Hambye:2019dwd} or footnote 8 in \cite{Cirelli:2016rnw}) and may have their own distinct temperature. We assume that the dark sector and the SM have a different temperature by introducing the dark-to-SM temperature ratio \cite{Cirelli:2016rnw}
\begin{equation}
\label{eq:tempratio}
\rtilde \equiv \frac{\tilde{T}_D}{\tilde{T_{SM}}},
\end{equation} 
where quantities with a $\sim$ on top are evaluated at some high temperature $\tilde{T}$.
Thus, the dark photon abundance before its decay is given by 
\begin{equation}
\label{eq:DP_comoving_number}
Y_{\mathsmaller{\rm V}} = \frac{n_{\mathsmaller{\rm V}}}{\sSM} = \frac{45 \zeta(3)}{2\pi^4} \frac{\gtildeD}{\gtildeSM} \rtilde^3 \simeq 0.0169 \left(\frac{\gtildeD}{6.5}\right) \rtilde^3 ,
\end{equation}
where $\gtildeD$ and $\gtildeSM$ are the relativistic number of degrees of freedom in the dark sector and the SM at temperature $\tilde{T}$. Plugging Eq.~\eqref{eq:DP_comoving_number} into Eq.~\eqref{eq:Tdom_def} implies a simple relation between the temperature at which the dark photon dominates the universe $T_{\rm dom}$ and its mass $\mV$. 
We choose to be agnostic about the mechanism setting the abundances in the dark sector and we enclose all possibilities by introducing a dark-to-SM temperature ratio $\tilde{r}$\footnote{
Production of the dark photon in the early universe has been studied in the literature. For
a small kinetic mixing $\epsilon$, the abundance of the dark sector can be set non-thermally
either by freeze-in \cite{Hall:2009bx, Chu:2011be, Berger:2016vxi, Hambye:2019dwd}, or by a separate reheating mechanism. In the latter case, the temperature asymmetry in Eq.~\eqref{eq:tempratio} results from an asymmetric reheating \cite{ Hodges:1993yb,Berezhiani:1995am,Feng:2008mu,Adshead:2016xxj}. For moderate kinetic mixing $\epsilon \gtrsim 10^{-6} \sqrt{\MDM/\text{TeV}}$ \cite{Hambye:2019dwd}, the dark sector may have been at thermal equilibrium with the SM, but asymmetric temperatures can result from asymmetric changes in relativisitic degrees of freedom \cite{Cirelli:2016rnw}. On the other hand, a possibility for thermally equilibrating the $U(1)_D$ sector and the SM in the case  of a small kinetic mixing $\epsilon$ would be to introduce a dark Higgs $\phi$, mixing with the SM Higgs, which once at thermal equilibrium with SM, decays into dark photons.}

\begin{figure}[h!]
\centering
\raisebox{0cm}{\makebox{\includegraphics[height=0.5\textwidth, scale=1]{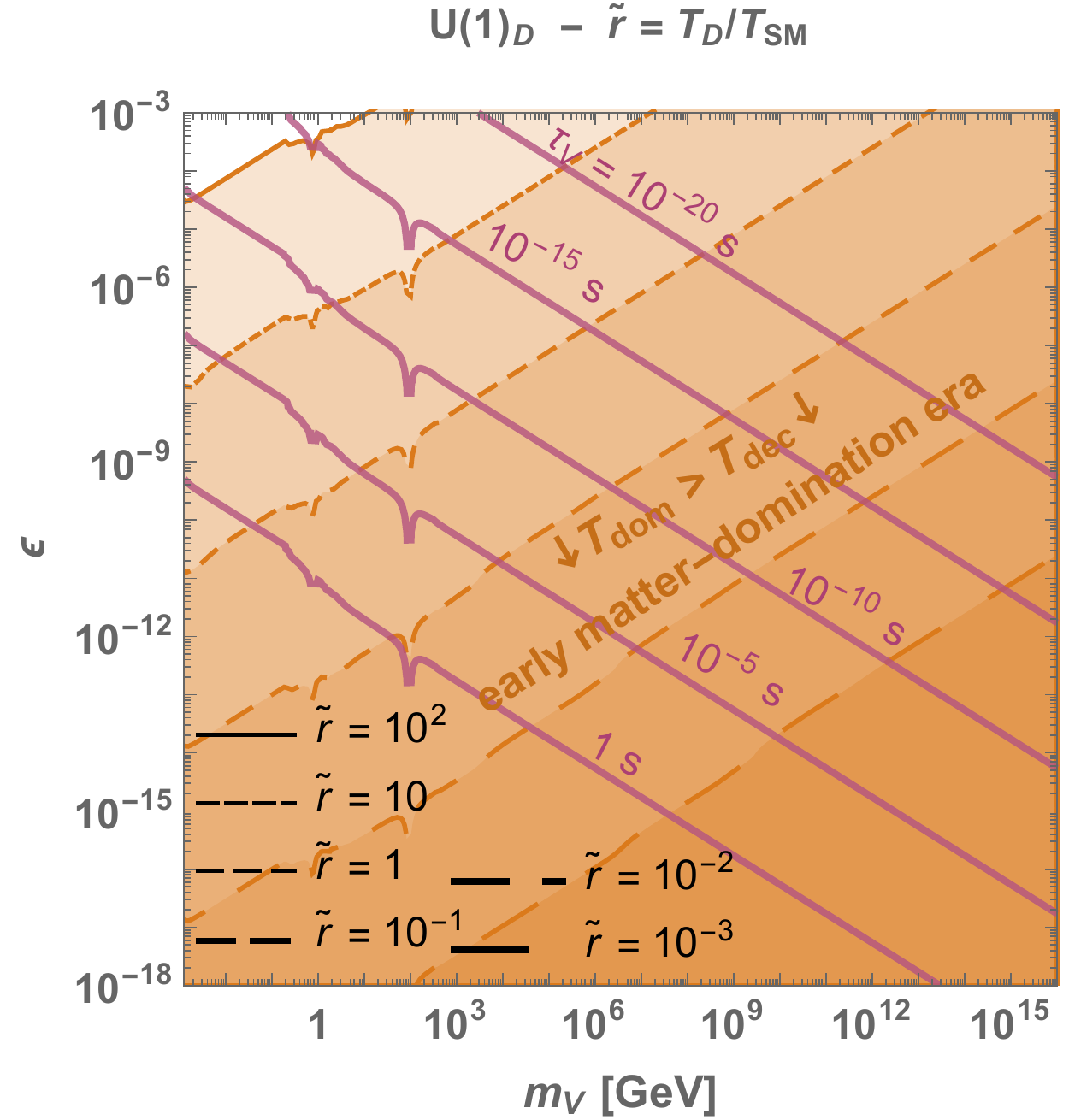}}}
\raisebox{0cm}{\makebox{\includegraphics[height=0.5\textwidth, scale=1]{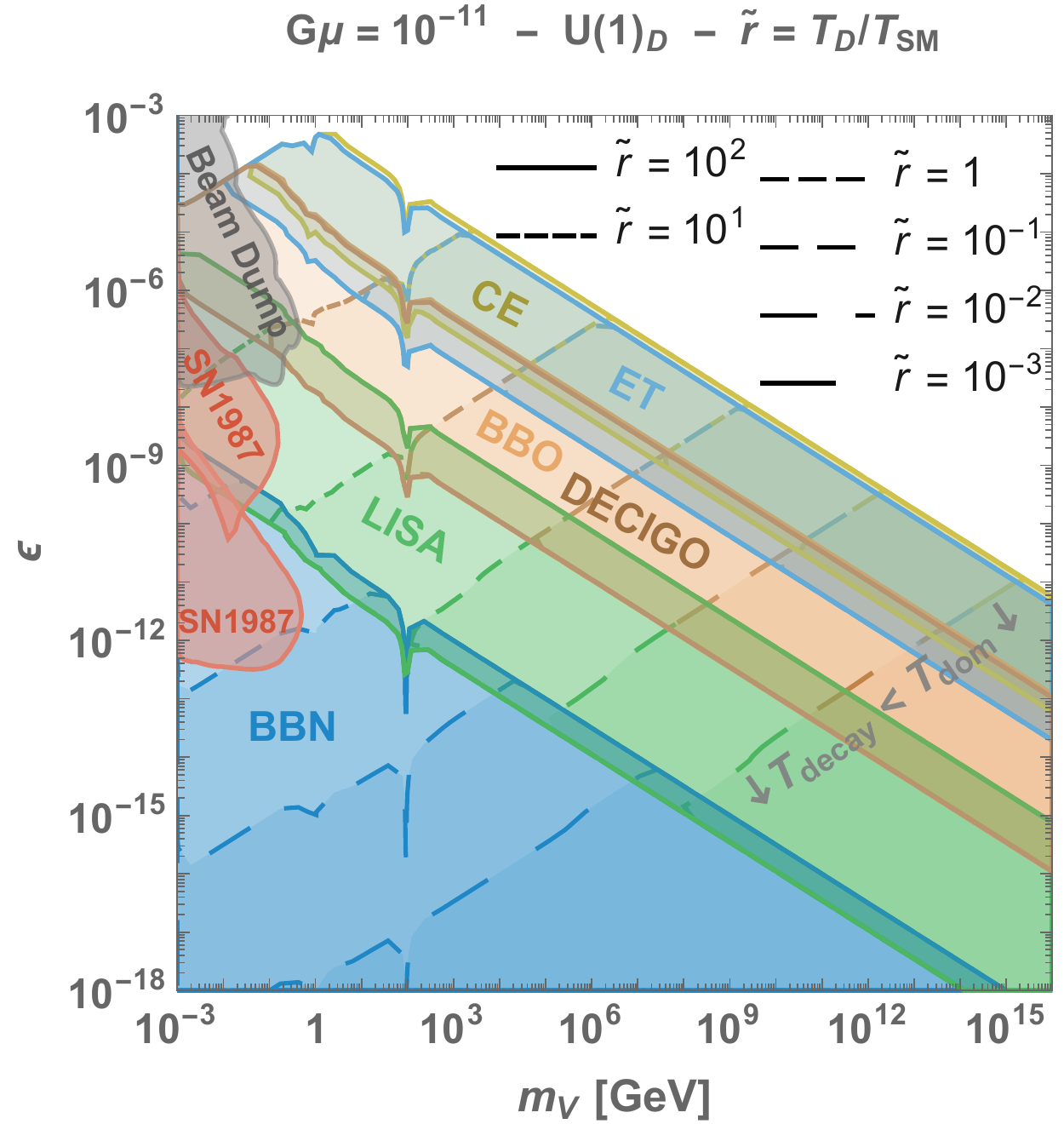}}}
\caption{\it \small \label{fig:GWconst_DarkPhoton1} \textbf{Left}: Constant dark photon lifetime $\tau_{V}$ contours. For a given dark-to-SM temperature ratio $\tilde{r}\equiv \TD/\TSM$, a non-standard early matter domination is induced below the corresponding orange line where the dark photon dominates the universe before it decays.
\textbf{Right}: Expected constraints on the dark photon mass $\mV$ and kinetic mixing $\epsilon$, assuming the measurement of a GW spectrum from CS with tension $G\mu = 10^{-11}$ by future GW interferometers. We use the \textit{turning-point prescription (Rx 1)} discussed in Sec.~\ref{sec:triggerMatterGW}.}
\end{figure}

As shown in the left panel of Fig.~\ref{fig:GWconst_DarkPhoton1}, low kinetic mixing $\epsilon$, large mass $\mV$ or large dark-to-SM temperature ratio $\tilde{r}$ lead to an early-matter-dominated era, triggered when $T_{\rm dom} \gtrsim T_{\rm dec}$. The non-detection with a future GW interferometer, of the imprint left by such a matter era, in the GW spectrum from CS, would exclude the existence of the dark photon for given values of the kinetic mixing, the dark photon mass and the dark-to-SM temperature ratio $(\epsilon, \, \mV, \, \tilde{r})$.  We show the GW-from-CS constraints on the dark photon in the right panel of Fig.~\ref{fig:GWconst_DarkPhoton1}, together with existing constraints coming from supernova SN1987 \cite{Kazanas:2014mca, Chang:2016ntp} and beam-dump experiments \cite{Cirelli:2016rnw}. Other constraints on lighter dark photons do not appear on the plot and are summarized in the reviews \cite{Jaeckel:2010ni,Essig:2013lka,Alexander:2016aln}.
We also include the BBN constraint which imposes the dark photon to decay before $\tauV \lesssim 0.1$~s \cite{Jedamzik:2006xz, Jedamzik:2009uy, Kawasaki:2017bqm} or later if the energy density fraction carried by the dark photon is smaller than $\sim 10\%$ \cite{Cirelli:2016rnw}.
Note, that only the BBN and the GW-from-CS constraints depend on the dark-to-SM temperature ratio $\tilde{r}$ which fixes the abundance of the dark photon in the early universe.

\begin{figure}[h!]
\centering
\raisebox{0cm}{\makebox{\includegraphics[height=0.48\textwidth, scale=1]{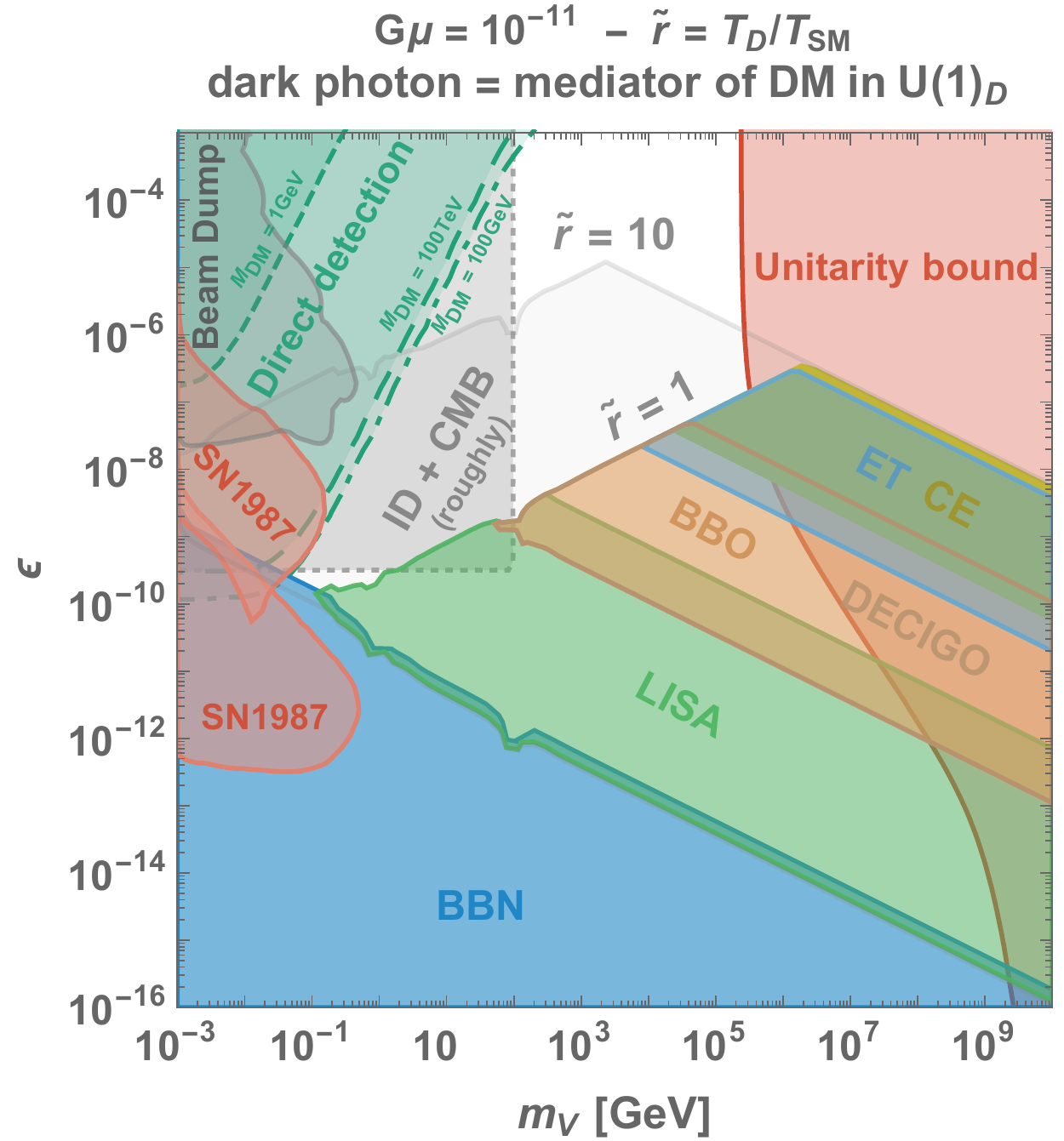}}}
\caption{\it \small \label{fig:GWconst_DarkPhoton2} 
Additional constraints when the dark photon is embedded in a DM model as the mediator of $U(1)_{D}$-charged DM (see text).
We compare the expected GW constraints from cosmic strings with the existing constraints on the $U(1)_{\rm D}$ DM model: Supernovae bounds from \cite{Kazanas:2014mca} and \cite{Chang:2016ntp}, direct detection bounds from \cite{Cirelli:2016rnw} and the indirect detection + CMB constraints are a rough estimate from \cite{Cirelli:2018iax}. Beam dump constraints are also taken from~\cite{Cirelli:2016rnw}. The unitarity bound on the DM mass $\MDM$ \cite{Griest:1989wd} can also be applied on the mediator mass because of the kinematic condition $\mV < \MDM$. The unitarity bound gets relaxed at small $\epsilon$ because of the larger entropy injection following the dark photon decay \cite{Cirelli:2018iax}. }
\end{figure}
We can appreciate the complementarity between the well-established supernova, beam dump, BBN constraints, and the expected constraints assuming the detection of a SGWB from CS by the GW interferometers. Indeed, whereas supernova and beam dump do not really constrain above $\mV \gtrsim 0.1$~GeV, the detection of a SGWB from CS with a string tension $G\mu \simeq 10^{-11}$ would exclude dark photon masses up to the maximal reheating temperature $\mV \sim 10^{16}$~GeV allowed by the maximal inflation scale $H_{\rm inf}\lesssim 6 \times 10^{13}$~GeV \cite{Ade:2018gkx, Akrami:2018odb}, and kinetic mixing as low as $\epsilon \sim 10^{-18}$.

\paragraph{The dark photon as a dark matter mediator:}
An interesting motivation for the dark photon is that it can play the role of a dark matter mediator. We can suppose that the dark sector also contains a Dirac fermion $\chi_{\mathsmaller{\rm D}}$ charged under $U(1)_{\mathsmaller{\rm D}}$, playing the role of DM \cite{Kors:2004dx, Feldman:2006wd, Fayet:2007ua, Ackerman:mha, Goodsell:2009xc, Morrissey:2009ur, Andreas:2011in, Goodsell:2011wn, Fayet:2016nyc, Cirelli:2016rnw, Cirelli:2018iax}
\begin{equation}
L \supset  \quad \bar{\chi}_{\mathsmaller{\rm D}}i\slashed{D}\chi_{\mathsmaller{\rm D}} - \MDM\bar{\chi}_{\mathsmaller{\rm D}}\chi_{\mathsmaller{\rm D}},
\label{lagrangian_U(1)_D_model}
\end{equation}
where $D_{\mu} = \partial_{\mu} + i g_{\mathsmaller{\rm D}} V_{\mathsmaller{\rm D}\mu}$ is the covariant derivative with $g_{\mathsmaller{\rm D}}$ the $U(1)_{\mathsmaller{\rm D}}$ gauge coupling constant. We suppose that the DM freezes-out by annihilating into pairs of dark photons,  we impose $\mV < \MDM$. We assume the dark photon to be non-relativistic when it decays but relativistic when it is produced, therefore, we set $\gtildeD = 3 + \frac{7}{8}\cdot 4 = 6.5$ in Eq.~\eqref{eq:DP_comoving_number}.

The unitarity bound on the DM mass $\MDM$ can be applied to the dark photon mass $\mV$ upon assuming $\mV < \MDM$. In the standard paradigm, the unitarity bound on s-wave annihilating dirac fermion DM is $\MDM \lesssim 140$~TeV \cite{Griest:1989wd, vonHarling:2014kha}. However, if long-lived and heavy, the decay of the mediator can, by injecting entropy, dilute the DM abundance and relax the unitarity bound to \cite{Cirelli:2018iax}
\begin{equation}
\MDM \lesssim 140~\text{TeV}~\sqrt{D},
\end{equation}
where $D$ is the dilution factor $D \simeq {T_{\rm dom}}/{T_{\rm dec}}$,
$T_{\rm dom}$ and $T_{\rm dec}$ are as defined in Eq.~\eqref{eq:Tdom_def} and Eq.~\eqref{eq:T_dec_tauX}.

In Fig.~\ref{fig:GWconst_DarkPhoton2}, we add the contraints on the dark photon when the later plays the role of the mediator of DM. They come  from direct detection \cite{Cirelli:2016rnw},  CMB \cite{Cirelli:2018iax},  indirect detection, using neutrino, gamma-rays, positrons-electrons and anti-protons \cite{Cirelli:2018iax}, as well as from unitarity \cite{Cirelli:2018iax}. 
They are complemented by the GW-from-CS constraints. For $\epsilon \lesssim 10^{-10}$, all the traditional indirect detection constraints evaporate and the unitarity bound is pushed to larger masses due to the entropy dilution following the dark photon decay such that the model is then currently only constrained by BBN. It is remarkable that  GW interferometers could probe this unconstrained region where $\epsilon<10^{-10}$ and $m_{V} > 1 $~GeV. In Fig.~\ref{fig:GWconst_DarkPhoton1} and ~\ref{fig:GWconst_DarkPhoton2}, we use the \textit{turning-point prescription (Rx 1)} discussed in Sec.~\ref{sec:triggerMatterGW}. Constraints are stronger when considering the \textit{spectral-index prescription (Rx 2)}, as shown in Fig.~\ref{fig:sketch} or in Fig.~\ref{fig:FullMD}.

\begin{figure}[h!]
\centering
\raisebox{0cm}{\makebox{\includegraphics[height=0.5\textwidth, scale=1]{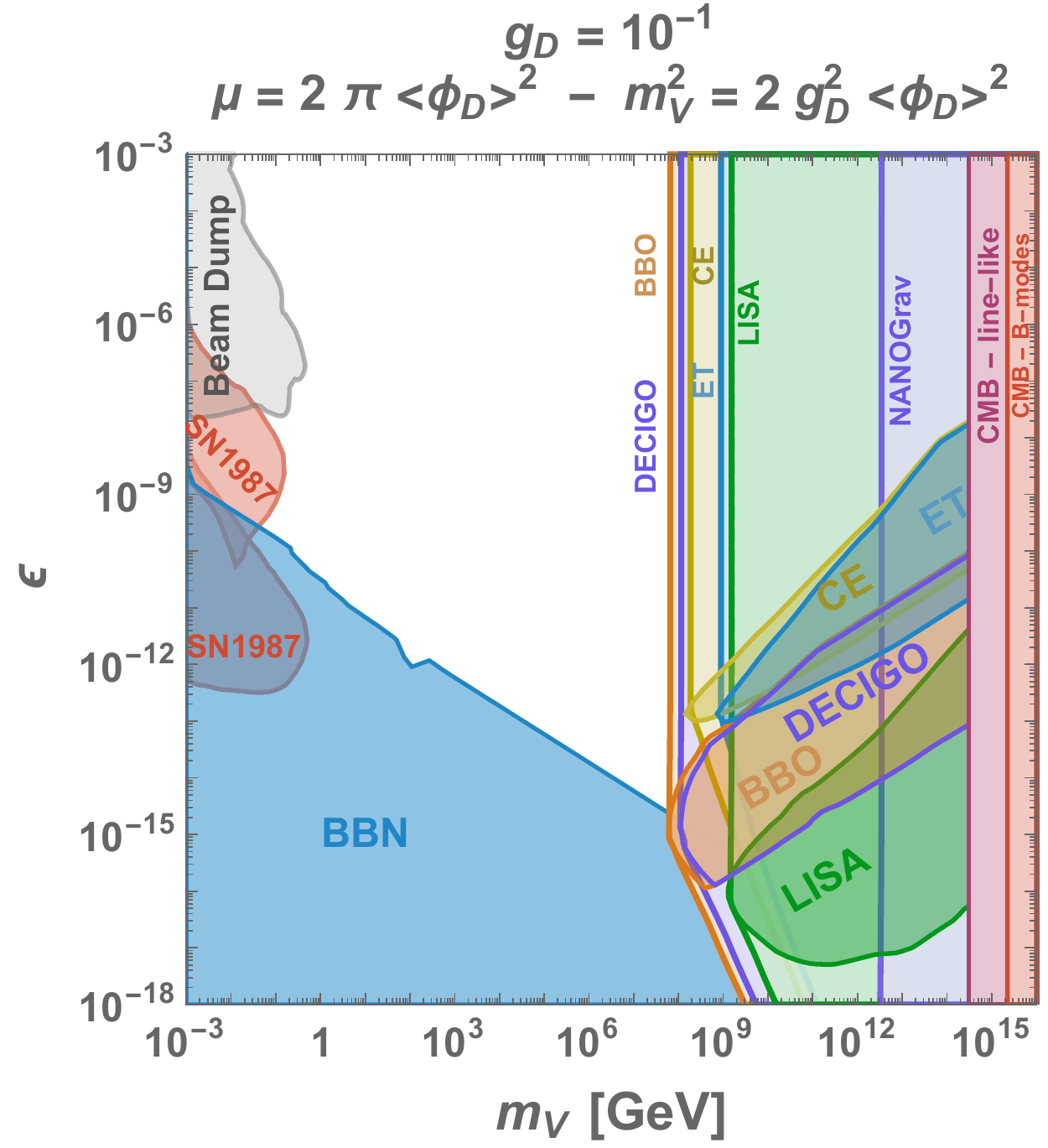}}}
\raisebox{0cm}{\makebox{\includegraphics[height=0.5\textwidth, scale=1]{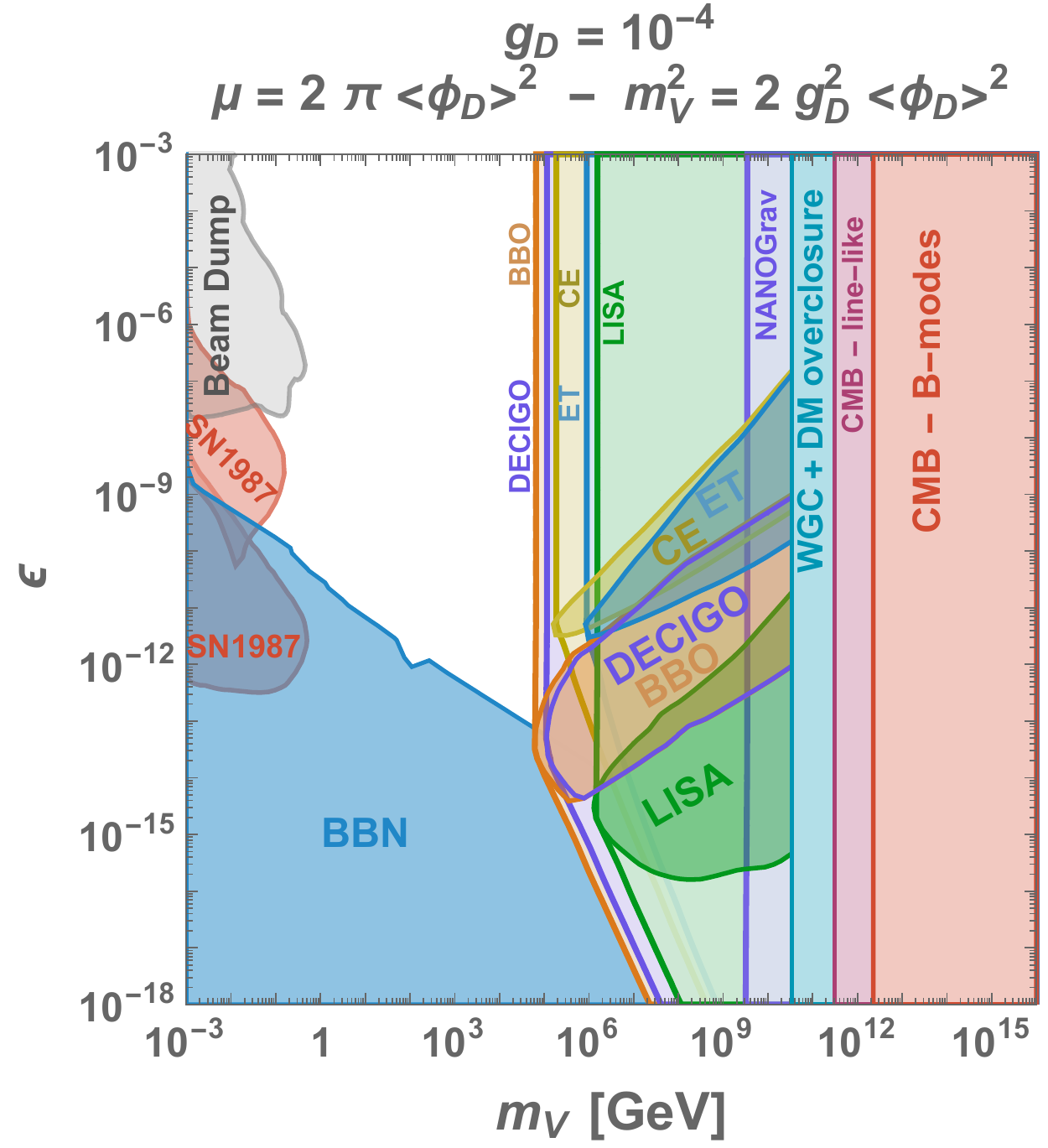}}}
\caption{\it \small \label{fig:U1_MD_gD} Scenario where the dark photon mass $\mV$ and the cosmic string network are generated by the spontaneous breaking of the same $U(1)$ symmetry, such that $\mV$ is related to the string tension $\mu$.  \textbf{Pale colors}: Constraints on the dark photon parameter space assuming the mere detection of the GW spectrum from CS by NANOGrav, LISA, ET, CE, DECIGO and BBO. \textbf{Opaque colors}: Constraints assuming the detection of the turning point in the GW spectrum induced by the transition from matter to radiation when the heavy dark photon decays. When combining Eq.~\eqref{eq:f_delta_RD} and Eq.~\eqref{eq:T_dec_tauX}, this last detection allows to measure the dark photon lifetime. The constraints described in the following part of this caption are independent of the GW emission. \textbf{Pale red}: The non-observation of the fundamental tensor B-modes in the CMB imposes the stringest upper bound on the energy density scale of inflation \cite{Akrami:2018odb}, $V_{\rm inf}  \lesssim 1.6 \times 10^{16}~\rm GeV$. This provides an upper-bound on the reheating temperature, which also must satisfy $\left< \phi_{\mathsmaller{\rm D}} \right> \lesssim T_{\rm reh}$ in order for the string network to be formed. Thus, we impose the CS formation to occur after the end of inflation with the following criteria: $\left< \phi_{\mathsmaller{\rm D}} \right> \lesssim V_{\rm inf}$.
\textbf{Pale purple}: Constraints from the non-observation of line-like temperature anisotropies in the CMB, e.g. \cite{Lizarraga:2016onn}, $G\mu \lesssim 2\times 10^{-7}$. \textbf{Pale sky blue}: In order to prevent DM overclosure, we must assume the $U(1)_D$ charged states to be heavier than the reheating temperature such that they are never produced. A possibility which is constrained by the Weak Gravity Conjecture, c.f. main text, thus we impose $\left< \phi_{\mathsmaller{\rm D}} \right> \lesssim g_{\mathsmaller{\rm D}}M_{\rm pl}$. The last inequality implicitly assumes $\left< \phi_{\mathsmaller{\rm D}} \right> \lesssim T_{\rm reh} $. Note however that such a charged state could be unstable, e.g. if it is a dark Higgs, in which case the WGC constraint is relaxed. }
\end{figure}

\paragraph{Scenario where the cosmic string network and the dark photon mass have the same origin:}

As a last remark, we comment on the case where the spontaneous breaking of the $U(1)_D$ symmetry would be responsible for the formation of the cosmic string network, so that the dark photon mass is no longer a free parameter but is related to the string tension $\mu$, through the Abelian-Higgs relations \cite{Vilenkin:2000jqa} 
\begin{align}
\label{eq:mu_phi}
\mu = 2\pi <\phi_{\mathsmaller{\rm D}}>, \\
\label{eq:mV_phi}
\mV^2 = 2 g_{\mathsmaller{\rm D}}^2 <\phi_{\mathsmaller{\rm D}}>^2,
\end{align}
where $\phi_{\mathsmaller{\rm D}}$ is the scalar field whose vacuum expectation value $<\phi_{\mathsmaller{\rm D}}>$ breaks the $U(1)_D$ symmetry spontaneously. 

In this case, we find that most of the relevant parameter space is ruled out due to overabundance of dark matter.\footnote{The cross-section of a pair of $U(1)_D$ fermions annihilating into dark photons is given by $\sigma v \simeq \pi \alpha_{\mathsmaller{\rm D}}^2/m_{\psi}^2$ with $\alpha_{\mathsmaller{\rm D}}=g_{\mathsmaller{\rm D}}^2/4\pi$. It is way too weak to prevent universe overclosure, except if we tune the Yukawa coupling of the fermion, $\lambda$, defined by $m_{\psi}=\lambda \left< \phi \right>/\sqrt{2}$, to very small values.} The only viable solution would be to assume that the states which are charged under $U(1)_D$ and stable under decay, are heavier than the reheating temperature such that they are never produced. 
The Weak Gravity Conjecture (WGC) requires the existence of a charged state with mass smaller than \cite{ArkaniHamed:2006dz}
\begin{equation}
m_{\rm X} \lesssim g_{\mathsmaller{\rm D}} M_{\rm pl}.
\end{equation}
Hence, $g_{\mathsmaller{\rm D}} M_{\rm pl}$ sets the maximal reheating temperature, above which charged states responsible for universe overclosure might be produced.
Therefore, we should exclude the parameter space where the temperature of the $U(1)_D$ spontaneous breaking, taken as $\sim\left< \phi_{\mathsmaller{\rm D}} \right>$, is heavier than $g_{\mathsmaller{\rm D}} M_{\rm pl}$, c.f. pale sky blue region in right plot of Fig.~\ref{fig:U1_MD_gD}.
Note that the WGC does not specify if the suggested charged state is stable under decay or not. For instance, it would be stable and overclose the universe if it is a $U(1)_D$ fermion but not if it is a $U(1)_D$ Higgs, which can still decay into a dark photon pair when $m_{\phi_{\mathsmaller{\rm D}}}\gtrsim 2\mV$. Hence, the WGC constraint in our parameter space has to be taken with a grain of salt.

Assuming a natural gauge coupling value, $g_{\mathsmaller{\rm D}}=10^{-1}$, we find that dark photons heavier than $\gtrsim 100~\rm PeV$ would be accompanied with a $U(1)_D$ cosmic string network producing an observable GW spectrum, see left plot of Fig.~\ref{fig:U1_MD_gD}.
In the case where $g_{\mathsmaller{\rm D}}=10^{-4}$, we could probe dark photon masses down to $\gtrsim 100$~TeV, see right plot of Fig.~\ref{fig:U1_MD_gD}.

On the same plot, we superpose the constraints, shown with pale colors, coming from the simple observation of the GW spectrum with future experiments (except NANOGrav which is already operating), and the constraints, shown with opaque colors, coming from the detection of the turning point where the spectral index of the GW spectrum changes due to the decay of the dark photon which was dominating the energy density of the universe.  The former detection would allow to measure the dark photon mass whereas the latter detection would allow to access its lifetime.

\section{Summary }

 If  future GW observatories have the sensitivity to detect stochastic GW backgrounds of primordial origin and to measure precise features in this spectrum, they can reveal very unique information about very high scale physics.
Particularly relevant sources of GW are cosmic strings. 
Cosmic strings are almost ubiquitous in many Grand-Unified Theories.  As they keep emitting throughout the whole cosmological history of the universe,  the resulting GW spectrum covers a wide range of frequencies and can be detected either by space-based or ground-based observatories.
An early era of matter domination due to new heavy particles generates a clear signature in the GW spectrum of cosmic strings.

In this study, we assume the existence of an early matter era due to the presence of a cold particle $X$ temporarily dominating the energy density of the universe and decaying before the onset of BBN. We compute its impact on the GW spectrum of CS beyond the scaling regime. 
We show that detecting such a feature and interpreting it in terms of a new heavy relic can lead to unparalleled constraints in the  $(\tau_X,\,m_XY_X)$ (lifetime, yield) plane.
In Fig.~\ref{fig:mY_tauX_GWI_VS_BBN}, we provide model-independent constraints which extend the usual BBN constraints on the lifetime $\tau_X$ by $15$ orders of magnitude for $G\mu=10^{-11}$, as we are able to constrain early matter dominated era ending when the temperature of the universe is between $50$~TeV and $1$~MeV.

Next, we show that this new search strategy is likely to provide unprecedented constraints on particle physics models. 
We illustrate this on  minimal models of massive particles.  In the first class, the heavy particle has only gravitational interactions and decays though a Planck-suppressed coupling, c.f. Fig.~\ref{fig:sketch}.
In the second class, the heavy relic is a new $U(1)$ gauge  boson that decays to the Standard Model via kinetic mixing to $U(1)_Y$ hypercharge.
We point out that supersymmetric  theories could be probed, well above the reach of present and future colliders, up to a gravitino mass scale of $10^{10}$~GeV, due to the presence of oscillating scalar moduli fields produced after dynamical supersymmetry breaking, c.f. Fig.~\ref{fig:GWconst_moduli}. Secondly, we study a simple model of massive scalar particle interacting only gravitationally  with the Standard Model and which therefore has no chance to be observed  in collider or direct/indirect detection experiments. If non-conformally coupled to gravity, it can be  abundantly produced by gravitational effects at the end of inflation, hence leading to a matter era,  which stochastic gravitational-wave backgrounds from cosmic strings can  uniquely probe, c.f. Fig.~\ref{fig:GWconst_OnlyGravCoupled}.
Finally, we study a model of dark photon kinematically coupled to the Standard Model hypercharge, possibly embedded in the $U(1)_D$ dark-photon-mediated dark matter model. The constraints we obtain from GW on $U(1)_D$ dark matter falls in the large mass/small kinetic mixing ballpark which is otherwise unreachable  by any current and probably future direct/indirect detection and CMB constraints, c.f. Fig.~\ref{fig:GWconst_DarkPhoton2}.  
{At last, we consider the possibility that  the dark photon mass and the cosmic string network are generated by the spontaneous breaking of the same $U(1)$ symmetry and show that we can use future GW interferometers to probe dark photon masses above $100 ~\rm PeV$, or even down to the $\rm TeV$ scale if we tune the gauge coupling to small values, see Fig.~\ref{fig:U1_MD_gD}.}

These are only a few minimal examples of particle physics models generating early matter eras. There are many other  well-motivated models which would deserve consideration in this respect.
We will present  the corresponding  constraints on axion-like-particles and primordial black holes in a separate study.

\medskip

\section*{Acknowledgements}

We thank Quentin Bonnefoy, Roberto Contino, Valerie Domcke, Yohei Ema, Filippo Sala and J\'er\^{o}me Vandecasteele for useful comments and discussions. 
This work is supported by the Deutsche Forschungsgemeinschaft under Germany's Excellence Strategy - EXC 2121 ``Quantum Universe" - 390833306. 
The work of Y.G. is partly supported by a PIER Seed Project funding `Dark Matter at 10 TeV and beyond, a new goal for cosmic-ray experiments' (Project ID PIF-2017-72).
P.S. acknowledges his master-degree scholarship from the Development and Promotion of Science and Technology Talents project (DPST), Thailand.

{\footnotesize
\bibliographystyle{JHEP}
\bibliography{ProbeHeavyRelicCS}
}

\end{document}